\documentclass[12pt,epsf]{article}
\usepackage{amssymb,amsmath,amsbsy}
\usepackage[dvipsnames]{xcolor}
\usepackage{graphicx,color}
\definecolor{purple}{rgb}{0.7,0.0,0.5}
\usepackage[bookmarks = false, colorlinks = true, linkcolor = black, citecolor = black, urlcolor=black]{hyperref}
\newcommand{\mT}{\mathcal{T}}

\newcommand{\ben}{\begin{enumerate}}
\newcommand{\een}{\end{enumerate}}
\newcommand{\be}{\begin{equation}}
\newcommand{\ee}{\end{equation}}
\newcommand{\bea}{\begin{eqnarray}}
\newcommand{\eea}{\end{eqnarray}}
\newcommand{\lb}{\left(}
\newcommand{\rb}{\right)}
\newcommand{\la}{\langle}
\newcommand{\ep}{\epsilon}
\newcommand{\ra}{\rangle}
\newcommand{\nn}{\nonumber}
\newcommand{\p}{\partial}
\newcommand{\mO}{{\mathcal O}}
\newcommand{\cR}{{\mathcal R}}

\newcommand{\nbox}{{\,\lower0.9pt\vbox{\hrule \hbox{\vrule height 0.2 cm \hskip 0.19 cm \vrule height 0.2 cm}\hrule}\,}}
\newcommand{\Tr}{\ {\rm Tr}\ }

\newcommand\ket[1]{\ensuremath{\lvert{#1}\rangle}}
\newcommand\bra[1]{\ensuremath{\langle{#1}\rvert}}
\newcommand\vev[1]{{\ensuremath{\left\langle{#1}\right\rangle}}}

\DeclareMathOperator{\csch}{csch}

\def\href#1#2{#2}

\newcommand\al{{\alpha}}
\newcommand\sig{\sigma}

\newcommand\lam{\lambda}

\newcommand\om{\omega}

\newcommand\De{{\ensuremath{{\Delta}}}}

\def\ov{\over}

\begin{document}
\begin{titlepage}
\hfill MIT-CTP/4956\\
\vbox{
    \halign{#\hfil         \cr
           } 
      }  
\vspace*{15mm}
\begin{center}

{\Large \bf 
Universality of Quantum Information in Chaotic CFTs}

\vspace*{13mm}
Nima Lashkari$^{a}$, Anatoly Dymarsky$^{b}$, and Hong Liu$^{a}$\\
\vspace*{10mm}
\let\thefootnote\relax\footnote{$\mathrm{lashkari@mit.edu},\  \mathrm{a.dymarsky@uky.edu},\  \mathrm{hong\_liu@mit.edu}$}
{${}^a$ Center for Theoretical Physics, Massachusetts Institute of Technology\\
77 Massachusetts Avenue, Cambridge, MA 02139, USA \\ \vspace*{0.2cm}
${}^b$ Department of Physics and Astronomy, University of Kentucky,\\
Lexington, KY 40506, USA \\
Skolkovo Institute of Science and Technology, Skolkovo Innovation Center, \\ Moscow 143026 Russia
}

\vspace*{0.7cm}
\end{center}
\begin{abstract}

We study the Eigenstate Thermalization Hypothesis (ETH) in chaotic conformal field theories (CFTs) of arbitrary dimensions. 
Assuming local ETH, we compute the reduced density matrix of a ball-shaped subsystem of finite size in the infinite volume limit when the full system is an energy eigenstate. This reduced density matrix is close in trace distance to a density matrix, to which we refer as 
the {\it ETH density matrix}, that is independent of all the details of an eigenstate except its energy and charges under global symmetries.  In two dimensions, the ETH density matrix is universal for all theories with the same value of central charge. 
We argue that the ETH density matrix is close in trace distance to the reduced density matrix of the (micro)canonical ensemble. 
We support the argument in higher dimensions by comparing the Von Neumann entropy of the ETH density matrix with the entropy of a black hole in holographic systems in the low temperature limit. Finally, we generalize our analysis to the coherent states with energy density that varies slowly  in space, and show that locally such states are well described by the ETH density matrix.

\end{abstract}

\end{titlepage}

\vskip 1cm

\section{Introduction and outline}

Quantum information plays an increasingly important role in our understanding
and characterization of quantum matter. The holographic duality together with the black hole information loss paradox give strong hints that quantum information is also likely to play a central role in our understanding of quantum gravity and the emergence of spacetime.

In this paper, we discuss the quantum information properties of {\it  chaotic} conformal field theories (CFTs) 
expanding on the observations made in an earlier paper~\cite{Lashkari:2016vgj}. We provide evidence that the quantum information content of highly excited energy eigenstates of  in conformal theories exhibit a great degree of universality.

We {\it define} chaotic quantum field theories (QFT) to be those satisfying a local version of the Eigenstate Thermalization Hypothesis (ETH)~\cite{Lashkari:2016vgj} (see \cite{srednicki1994chaos,Srednicki:1995pt} for ETH in generic quantum systems including density matrix formulation \cite{garrison2015does,Dymarsky:2016aqv}). More explicitly,  we say that a QFT on a homogenous compact space satisfies local ETH if for a  local operator $\mO_p$ (with $p$ labeling different operators),
\bea\label{ETH}
\la E_a|\mO_p|E_b\ra= O_p (E)\delta_{ab}+ \De_{p ab},
\eea
where $\ket{E_a}$ is a highly excited energy eigenstate, the diagonal element $O_p (E)$ is a smooth function of $E = {E_a + E_b \over 2}$, and  $\De_{p ab} \sim  e^{- O(S(E))}$  where $e^{S (E)}$ is the density of states at energy $E$. If $\ket{E_a}$ has other quantum numbers associated with other global symmetries, $O_p (E)$ can also  smoothly depend on those quantum numbers. To simplify the notation, we will suppress such dependence. In case of CFTs, definition of ETH  \eqref{ETH} will require additional clarifications which we explicitly described below.

The high-energy eigenstates of a quantum many-body system are, in general, hard to access, and until now essentially all discussions of ETH have been 
limited to direct numerical diagonalizations (for instance see \cite{rigol2008thermalization}). With the current computational resources, a direct numerical diagonalization approach to QFT seems unrealistic.  In~\cite{Lashkari:2016vgj}, we advocated that CFTs provide an exciting laboratory for exploring the implications of ETH and {potentially even} proving it. 
In a CFT,  due to the state-operator correspondence, the energy eigenstates can be represented as local operators with definite scaling dimensions, and~\eqref{ETH} becomes a condition on the operator product expansion (OPE) coefficients. 
This opens up many powerful analytic tools for studying ETH. The previous studies of ETH in CFTs that have been  inspiration for our work are \cite{Asplund:2014coa,Fitzpatrick:2015zha,Fitzpatrick:2015qma,He:2017vyf,Basu:2017kzo,He:2017txy}.

More explicitly,  consider a $(d+1)$-dimensional CFT on a $d$-dimensional sphere  $\mathbb{S}^d$ with radius $L$. 
Since a primary operator and its descendants are algebraically related, the equation~\eqref{ETH} written for CFTs should restrict only to the states $| E_a\ra$ dual to primary operators \cite{Lashkari:2016vgj}. In particular, for two-dimensional CFTs, $\ket{E_a}$ should correspond to Virasoro primary operators.\footnote{In every two-dimensional CFT there is an infinite number of conserved charges associated to the KdV hierarchy \cite{bazhanov1996integrable}. As we will discuss later, for a Virasoro primary, all these charges are fixed in terms of the conformal dimension, therefore $O_p (E)$ depends only on $E$.}
Without loss of generality, we further restrict to scalar primary operators $\Psi_a$ of dimension $h_a = E_a L$. The energy density of the system in such a state is 
\be 
\ep_a = {E_a \over L^d \omega_d} = {h_a \over L^{d+1} \omega_d},
\ee
where $\omega_d$ is the volume of a unit sphere $\mathbb{S}^d$. For a CFT in a 
 thermal state of temperature $T$, $\ep_a \sim d_T  T^{d+1}$ where $d_T$ is the normalization of the two-point function of stress tensor (\ref{twopointstress}). 
This motivates us to define the ``thermal'' length scale associated with $\ket{E_a}$ as
 \be \label{uj}
 \lambda_T=\left({\ep_a \over d_T}\right)^{-{1 \over d+1}}  \sim T^{-1} 
 \ .
\ee

In the thermodynamic limit with $L \to \infty$, while keeping energy density $\ep_a$ finite, and hence a finite 
$\lambda_T$,  the scaling dimension $h_a$ scales with $L$ as 
\be\label{1ex}
h_a = d_T \omega_d \left({L \over \lambda_T}\right)^{d+1}\ .
\ee 
Applying the conformal transformation that maps the cylinder $\mathbb{S}^d\times \mathbb{R}_t$ to $\mathbb{R}^{d+1}$ (the radial quantization frame) the local ETH condition~\eqref{ETH} translates into a statement about the OPE coefficient $C^p_{ab}$ multiplying the operator $\mO_p$ appearing in the expansion of two primaries $\Psi_a$ and $ \Psi_b^\dagger$ corresponding to $\ket{E_a}$ and $\bra{E_b}$,
\bea\label{ope}
&&C_{pab}L^{-h_p} = O_p (E)\delta_{ab}+ \De_{pab}\ , \nn\\
&&\la \Psi^\dagger_b(\infty)\mO_p(1)\Psi_a(0)\ra=C_{pab}\ ,\nn\\
&&\Psi_a\times\Psi_b^\dagger=\sum_p C^p_{ab}\mO_p\ .
\eea
We raise and lower the $p$ index of $C^p_{ab}$ using the Zamolodchikov metric $\la \mO_p(1) \mO_p(0)\ra=d_p$.
In the thermodynamic limit, under the assumption that~\eqref{ETH} {applies} for any operator $\mO_p$ of dimension $h_p$, which we keep fixed as $L$ becomes large, the equation~\eqref{ope} implies that  the OPE coefficient $C_{ab}^p$ must scale with $h_a \to \infty$ as 
\be \label{eo}
C_{p ab} 
= h_a^{h_p\over d+1} (d_T \omega_d)^{- {h_p\over d+1}}  \delta_{ab} f_p (E) + \cR_{p ab}\ . 
\ee
Here, the correction term $\cR_{p ab} = L^{h_p} \Delta_{p ab}\sim e^{- O(h_a^{d \over d+1}) + {h_p \over d+1} \log h_a} $ 
is exponentially small in $h_a$,  and $f_p (E) =  \lambda_T^{h_p}  O_p (E) $ is a smooth dimensionless function of $E$.  Since there are no other dimensionfull parameters in the problem,  $f_p (E)$  then has to be a constant, independent of $E$, i.e. 
\be \label{eo1}
C_{p ab} 
= h_a^{h_p\over d+1} (d_T \omega_d)^{- {h_p\over d+1}}  \delta_{ab} f_p + \cR_{p ab}\ . 
\ee
We stress that the equation~\eqref{eo1} encodes the following nontrivial implications of the local ETH. (i)  Operators $\mO_p$ whose $C_{p aa}$ grow slower than $h_a^{h_p\over d+1}$ with $h_a$ cannot have a non-vanishing expectation value in the thermodynamic limit, while it is impossible {for the OPE coefficient} $C_{ab}^p$ to grow faster than $h_a^{h_p\over d+1}$ as that would imply thermodynamic limit for such a theory does not exist.  (ii) The spectrum of  operators $\mO_p$ appearing in the OPE of $\Psi_a$ and $\Psi^\dagger_a$ is independent of specific properties of $\Psi_a$, and only depends on its scaling dimension (energy).

Integrable systems are expected not to satisfy the local ETH. A simple example is a two-dimensional free massless boson on a spatial circle. This theory has heavy coherent primary states $e^{i\alpha \phi}|\Omega\ra$ with large dimension $h_{\alpha}=|\alpha|^2/2\gg 1$. The OPE coefficient of this heavy state with a primary of dimension one, $\p\phi$, explicitly violates~\eqref{eo} since it grows as 
\be 
C_{e^{i\alpha \phi},e^{-i\alpha \phi}}^{\p\phi}\sim \alpha\sim \sqrt{h}_\alpha ,
\ee
while from the thermal expectation value of $\p\phi$ we know that $f_{\p\phi}$ on the right-hand-side of (\ref{eo}) is zero.

\begin{figure}
\centering
\includegraphics[width=0.8\textwidth]{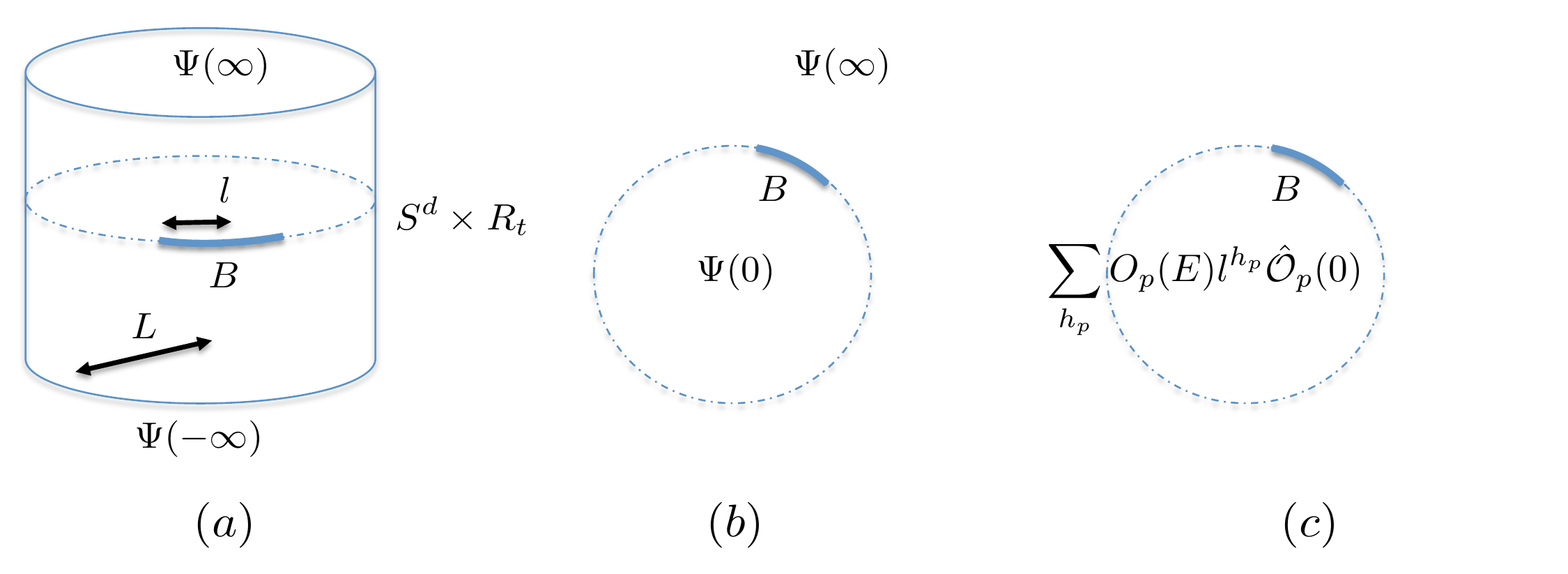}\\
\caption{\small{ (a) The cylinder $\mathbb{S}^d\times \mathbb{R}_t$ frame and the Euclidean path-integral that prepares the the density matrix in the eigenstate corresponding to $\Psi$ on subsystem $B$ (b) The same path-integral in the radial quantization $\mathbb{R}^{d+1}$ conformal frame (c) The path-integral for $\psi_{ETH}$ in the radial quantization frame.}}
\label{fig2p}
\end{figure}

Now consider a {chaotic} CFT  in a highly excited energy eigenstate. We focus on the reduced density matrix of a ball-shaped region $B$ of size $l$ inside $\mathbb{S}^d$ of size $L$  and consider the thermodynamic limit $L\rightarrow \infty$ with $l$ kept fixed. {The complement of $B$ inside $\mathbb{S}^d$ will be denoted as $B^c$.}
It was shown in~\cite{Lashkari:2016vgj} that the reduced density matrix $\psi_a(B)  \equiv {\rm Tr}_{B^c} \ket{E_a} \bra{E_a}$ {for the system in} state $\ket{E_a}$ can be well approximated by a density matrix $\psi_{\rm ETH}(B,E)$, to which we will refer as an {\it ETH density matrix}. $\psi_{ETH}$ depends only  on $B$ and energy $E_{a}$
\be 
 \label{cmeth}
||\psi_a(B) - \psi_{\rm ETH}(B,E = E_a) ||   \sim e^{- O(S(E_a))}\ , 
\ee
where  $\|\cdots\|$ is the trace distance. In particular, it was shown that the ETH density matrix $\psi_{\rm ETH}(B,E)$ can be written as 
\be\label{uni}
\psi_{\rm ETH}(B,E) = \sum_{h_p} O_p(E) l^{h_p} \hat{\mO}_p(0),  \qquad  \hat{\mO}_p=U^\dagger \mO_p U,
\ee
where ${\mO}_p$ denotes the family of operators which appear in the OPE of $\Psi_a$ and $\Psi^\dagger_a$, $O_p (E)$ {denotes their expectation values}~\eqref{ETH}, and  $U$ is the unitary operator corresponding to the conformal transformation from the Rindler frame to the radial quantization frame; see figure \ref{fig2p}(c).  Equation~\eqref{uni} defines a density matrix on $B$ as being prepared via a Euclidean path-integral over $\mathbb{R}^{d+1}$ with the specified boundary conditions ``above" and ``below" $B$ within $\mathbb{S}^d$ of unit radius, and the sum of local operators on the right hand side of~\eqref{uni} inserted at the origin of $\mathbb{R}^{d+1}$ (see figure \ref{fig2p}). 
We will see later that the domain of convergence of this sum is fixed by the conformal symmetry to be infinite.

Expressing $O_p (E)$ in terms of constants $f_p$ of~\eqref{eo1}, we find that \eqref{uni} is an expansion in ${l \over \lam_T}$
\be\label{uni1}
\psi_{\rm ETH}(B,E)=\sum_{p } f_p \left({l \over \lam_T}\right)^{h_p} \hat{\mO}_p(0) \ .
\ee
 In the low temperature regime ${l \over \lam_T} \ll 1$, it is enough to keep the first few terms while in the high temperature limit ${l \over \lam_T} \to \infty$ one has to sum the whole series, which should be convergent  for any large but finite $l/\lambda_T$.

In this paper, we first give a general argument that the ETH density matrix~\eqref{uni1} is close in trace distance 
to the reduced density matrix of a thermal state (there are subtleties in 2d).  Thus, by denoting the set of primary (quasi-primary in 2d) operators of a CFT that have non-zero thermal one-point functions by $\mathcal{A}_{therm}$, we can also write~\eqref{uni1} as 
\be\label{uni2}
\psi_{\rm ETH}(B,E)=\sum_{p \in \mathcal{A}_{therm}} f_p \left({l \over \lam_T}\right)^{h_p} \hat{\mO}_p(0) \ .
\ee
All (quasi-)primary operators that are not in $\mathcal{A}_{therm}$, and all the descendant fields drop out  in the thermodynamic limit from the sum \eqref{uni2}.   
We then discuss in detail the structure of the expansion~\eqref{uni1} in the low temperature regime. 

Note that the reduced density matrix in the eigenstate is close to the ETH density matrix (\ref{uni1}) (before we discard descendant fields) with exponential precision in $S(E)$, as dictated by local ETH. However, the convergence of the ETH density matrix to the reduced thermal state is controlled with corrections that are polynomially supressed in $S(E)$, as is the case anytime we compare quantities calculated in the microcanonical and the canonical ensembles.

\ben 

\item   In two dimensions ($d=1$),  the only Virasoro primary operator which has non-zero thermal value is the identity operator. Therefore, the ETH density matrix $\psi_{\rm ETH}(B,E)$ of~\eqref{uni} is solely expressed in terms of the Virasoro descendants of identity, i.e. $O_p (E)$ that are the polynomials of stress tensors and their derivatives. 
All $f_p$'s that correpond to the quasi-primaries in the Virasoro indentity block are fixed by the Virasoro algebra, and hence are independent of any specific properties of the 2d CFT except for the value of the central charge.
The ETH density matrix in 2d is universal across all CFTs with the given value of central charge, thus we refer to it as the {\it universal density matrix}.
We argue that if~\eqref{ETH} holds for Virasoro primaries, the subsystem density matrix in the eigenstate is well approximated by the universal density matrix.
Furthermore, we argue that the universal density matrix in the thermodynamic limit is close to the
reduced Generalized Gibbs Ensemble (GGE) provided we can map all their conserved charges.  That is to say
\be \label{gge}
\psi_{\rm univ}={1 \over Z} {\rm tr}_{B^c}\lb e^{-\beta H+\sum_i \mu_i Q_i} \rb+O(1/\sqrt{L}),
\ee
where the inverse temperature $\beta$ and the charges $\mu_i$ are chosen such that the GGE has the same {value of $Q_i$} charges as the universal density matrix. The conserved charges $Q_i$ are the infinite set of Korteweg-de Vries integrals of motions in two-dimensional CFTs \cite{bazhanov1996integrable}. Due to the complexity of evaluating the expectation values of $Q_i$
in the GGE, we are not able to provide a direct support for~\eqref{gge} at this point. Note that 
CFT formulation of ETH does not require (\ref{gge}) to hold. The equation (\ref{gge}) should hold if we further assume that one can solve for $\mu_i$ such that the GGE has the same values of charges $Q_i$ as the pure state.

In the limit that the central charge $c$ goes to $\infty$, we show that all the $\mu_i =0$ and 
the universal density matrix becomes close in trace distance to the standard Gibbs state. This is consistent with previous results of~\cite{Asplund:2014coa,Basu:2017kzo}.\footnote{As we explain in detail in section 3 equivalence of $\psi_{\rm ETH}$ and the reduced Gibbs state  does not imply that corresponding higher Renyi entropies for $n>1$ would have to match, and we find that they, indeed, do not match.}


\item In higher dimensional CFTs, in general, the polynomials of the stress tensor do not exist in the spectrum as primary operators. Furthermore, the conformal symmetry is a lot less restrictive than 2d, and any primary operator can have nonzero $O_p (E)$. 
It is natural to expect, and we provide further support in section \ref{sec:two}, that~\eqref{uni1} sums into the standard thermal ensemble
\be \label{therm}
\psi_{\rm ETH}={1 \over Z} {\rm tr}_{B^c}\lb e^{-\beta H} \rb+O(1/\sqrt{L})\ ,
\ee
where the inverse temperature $\beta$ is again chosen such that the thermal density matrix has the same energy $E$
 as the ETH density matrix. We provide support for~\eqref{therm} by
 computing the entanglement entropy of the ETH density matrix to the order $(l/\lambda_T)^{2(d+1)}$ and matching the answer with the holographic entanglement entropy of the same subsystem  as computed with the Ryu-Takayanagi formula in a black hole {background}. Note that up to this order, the entanglement entropy {exhibits universality and} depends only on the energy density and $d_T$, the two-point function of stress tensor. That is why one can match the answer with holography.
 

\een

The plan of the paper is as follows. 
In  Sec.~\ref{sec:nee} we give a general discussion of the relation between the ETH density matrix and that of a thermal state.  In Sec.~\ref{sec:2d} we discuss the structure of the ETH density matrix for a two dimensional CFT in detail. 
 In Sec.~\ref{sec:hd} we study the subsystem ETH in CFTs of dimensions larger than two. In Sec.~\ref{sec:gen} we consider states that have spatial and time dependence at scales much larger than the subsystem size and show that the {same} universal density matrix remains a good approximation {to describe local physics}.

\section{ETH density matrix and thermal states}\label{sec:nee}

We start with a brief discussion of various thermal ensembles 
for CFTs. The goal is to show that local ETH~\eqref{ETH} implies that the expectation values of 
$O_p$ in eigenstates as defined in~\eqref{ETH} coincide with the thermal averages. This enables us to show that 
the reduced density matrix of an energy eigenstate is close in trace distance to those of various thermal ensembles. 


\subsection{Different ensembles} 

Consider a QFT with a number of global symmetries living on a sphere. 
The microcanonical ensemble $\rho_{\rm micro}(E_0; \{\al\})$ 
is defined as an equal-weight average over all energy eigenstates lying within a narrow band around $E_0$ with a given set of quantum numbers $\{\al\}$ 
under various global symmetries, 
\be 
\rho_{\rm micro} (E_0; \{\al\}) = {1 \ov {\cal N}} \sum_{E \in (E_0- \De, E_0 + \De), {\rm given} \, \{\al\} }   \ket{E, \{\al\}} \bra{E, \{\al\}} \ .
\ee
As always, we choose the energy band width $\De$ to be much larger than the average level spacing that scales like $\exp(-O(L^d))$, but much smaller 
than the typical energy scales of interest. Here, ${\cal N}$ is the total number of states in the band. The density matrix of the canonical ensemble is 
\be 
\rho_{\rm can} (\beta, \{\al\}) = {1 \ov Z_{\{\al\}}}  e^{-\beta H}  P_{\{\al\}}, \qquad
Z_{\{\al\}} = \Tr P_{\{\al\}}  e^{-\beta H}
\ee
where $P_{\{\al\}}$ denotes projection into the subspace of the Hilbert space with given $\{\al\}$. 
The grand canonical density matrix is defined as 
\be 
\rho_{\rm grand} (\beta, \{\mu\}) = {1 \ov Z_{\{\mu\}}}  e^{-\beta H - \sum_i \mu_i Q_i} , \qquad 
Z_{\{\mu\}} = \Tr   e^{-\beta H- \sum_i \mu_i Q_i}
\ee
where $Q_i$ denote the complete set of commuting charges and $\{\mu\}$ denotes the collection of the
corresponding chemical potentials. 

For a general quantum field theory, in the thermodynamical limit, for a local operator $\mO$ whose quantum numbers we keep fixed as  the volume goes to infinity, the microcanonical,  canonical, and grand canonical  averages are all equivalent by the standard arguments, provided that one chooses $\beta$ and $\{\mu\}$ to give the average energy $E_0$ and the average charges $\{\al\}$. 
For example, the micro-canonical and the canonical ensemble which average over rotationally-invariant states (i.e. with $J^2=0$ where $J^2$ denotes the Casimir operator of the rotation group) are equivalent to the grand canonical ensemble with the corresponding $\mu_i =0$.  

The equivalence of ensembles in conformal field theory is more intricate since the representations of a conformal group are infinite dimensional. Furthermore, the states which lie in the same representation of the conformal group in general do not have the same energy. 
Let us first consider a CFT in $d > 2$. In this case, the conformal group is the higher dimensional Mobius transformations, and there are no new conserved charges beyond the generators of the conformal transformations. 
For  convenience, let us introduce $\hat \rho_{\rm micro}^{\rm (0)} (E_0; \{J^2 =0\})$ as the (un-normalized) microcanonical density matrix  of scalar primaries with energies in a narrow band around $E_0$,
where one sums over only the energy eigenstates which are scalar primaries.
 Similarly we can define $\hat \rho_{\rm micro}^{\rm (n)} (E_0; \{J^2=0\})$ to be the ensemble of states that descend at level $n$ from primary states of energy $E_0$.  
A state in the subspace defined by $\hat \rho_{\rm micro}^{\rm (n)} (E_0; \{J^2=0\})$ has energy approximately equal to $E_0 + O({n \ov L})$. 
The standard  microcanonical ensemble can then be expressed as 
\be \label{hnn}
\rho_{\rm micro} (E_0; \{J^2=0\})   = {1 \ov {\cal N}} \sum_{n}
\rho_{\rm micro}^{(n)} \left(E_0 - O\lb {n \ov L}\rb; \{J^2 =0 \}\right),
\ee
where ${\cal N}$ is the total number of states at energy $E_0$ including both primaries and descendants. 

Now, we consider the thermodynamic limit that is $L \to \infty$ with $E_0/L^{d}$ fixed. In this limit, from~\eqref{ETH} we have that for any $n$ which does not scale with $L$ 
\be 
\vev{E_0| \mO|E_0} = \vev{E_0 - O\lb {n \ov L}\rb | \mO|E_0 -O\lb {n \ov L}\rb}  + O(L^{-1}),
= \vev{E_0^{(n)} | \mO|E_0^{(n)}}  + O(L^{-1}) 
\ee
where $\ket{E_0}$ denotes a primary state while $\ket{E_0^{(n)}}$ denotes an $n$-th  
level descendant state of a primary state of approximate energy $E_0 -O({n \ov L})$; see \cite{Lashkari:2016vgj}. The density of states grows exponentially with energy
\bea
\log\Omega(E)\sim O(E^\alpha)\qquad 0<\alpha<1.\nn
\eea
The contribution of states in~\eqref{hnn} with $n$ scaling as $L$ or larger, is exponentially suppressed compared to the contribution of those with $n=0$; hence we neglect such states.
We conclude that in the thermodynamic limit for any local operator 
\be  \label{eeh}
\vev{E_0|\mO|E_0} = \Tr \left(\mO \rho_{\rm micro} (E_0; \{J^2 =0\}) \right)  + O(L^{-1}) \ 
\ee
and  will also be the same as in the canonical and grand canonical ensembles. 

A CFT in  $d=2$  
 has an infinite number of conserved charges that commute with both $L_0$ and $\bar{L}_0$. This is the KdV hierarchy of charges
$\{Q_{2k+1}, \bar Q_{2k+1}, \; k=1,2, \cdots\}$. Here, the corresponding microcanonical and canonical ensembles are denoted as
\be 
 \rho_{\rm micro} (E_0; \{Q_{2k+1}, \bar Q_{2k+1}\}) , \qquad 
  \rho_{\rm canonical} (\beta; \{Q_{2k+1}, \bar Q_{2k+1}\})
\ee
and the corresponding grand canonical ensemble is the so-called Generalized Gibbs Ensemble (GGE)
 \bea
 \rho_{GGE} (\beta; \{\mu_{2k+1}, \bar \mu_{2k+1}\}) =\frac{e^{-\beta(L_0+\bar{L}_0)-\sum_k \mu_{2k+1}Q_{2k+1}-\sum_k \bar{\mu}_{2k+1}\bar{Q}_{2k+1}}}{Z} \ .
 \eea
Again, $ \rho_{\rm micro} (E_0; \{Q_{2k+1}, \bar Q_{2k+1}\})$ contains descendant states. By descendants we are now referring to Virasoro descendants. Following the same arguments as above we conclude that 
\be
\vev{E_0,  \{Q_{2k+1}, \bar Q_{2k+1}\}|\mO|E_0,  \{Q_{2k+1}, \bar Q_{2k+1}\} } 
= \Tr \left(\mO \rho_{\rm micro} (E_0;  \{Q_{2k+1}, \bar Q_{2k+1}\}) \right) .
\ee
The same holds also for the canonical ensemble and the GGE,  provided we assume an appropriate growth of the density of states $\Omega$  as a function of $Q$.

\subsection{Equivalence of  reduced density matrices} 
\label{sec:two}

We now present a general argument showing that given~\eqref{eeh}, the reduced density matrix 
for a region $B$ of $\ket{E_0}\bra{E_0}$, and the ETH density matrix $\psi_{\rm ETH}$ are close in trace distance to
the reduced state $\rho$ of the subsystem $B$ of a thermal state (the two-dimensional case is different and will be discussed in more depth in section 4). The argument works for any of the three ensembles mentioned earlier.

The reduced density matrix of a region $B$ is a map from the observables living on $B$ to the expectation values. In conformal field theory, if $B$ is a topologically-trivial region the set of local operators on $B$ provide a basis for all operators in $B$.
One can compute the expectation value of a $k$-point function of operators local in the subsystem $B$ in a reduced state such as $\rho$ or $\psi_{\rm ETH}$ by successively applying OPEs to reduce the $k$-point function to a one-point function. This is possible because neither $\rho$ nor $\psi_{\rm ETH}$ have any operator insertions in their corresponding Euclidean path-integrals that limits the domain of the convergence of OPEs on the subsystem.

 
Consider  any two reduced density matrices $\rho$ and $\sigma$ whose Euclidean path-integral definitions do not involve  any operator insertions that limits the subsystem OPE. We will now show that $\rho=\sigma$ if and only if they have the same expectation value for all the local operators. The proof is a simple application of the Pinsker inequality:
 \bea\label{Pinskersym}
 \|\rho-\sigma\|^2\leq \frac{1}{2}\lb S(\rho\|\sigma)+S(\sigma\|\rho)\rb=\Tr((\rho-\sigma)(K_\sigma-K_\rho)) 
 \eea
 where $K_\rho$ and $K_\sig$ denote the modular operators for $\rho$ and $ \sig$, respectively.
 The modular operators of both $\rho$ and $\sigma$ can be expanded as
\bea
K=\sum_p  l^{h_p-(d-1)}\int_x f_p(x)\mO_p(x)+\sum_{p,q} l^{h_p+h_q-2(d-1)}\int_{x,y} f_{p,q}(x,y)\mO_p(x)\mO_q(y)+\cdots
\eea
where $p$ sums over the set of all local operators. We can use the OPEs of operators in conformal field theory to reduce the expression above to an infinite sum over local operators
\bea
K=\sum_p  l^{h_p-(d-1)}\int_x \, \tilde f_p(x)\mO_p(x) \ .
\eea
From (\ref{Pinskersym}) it then follows that if all the one-point functions of local operators match then the density matrices are the same. Now, imagine that the two density matrices have matching one-point functions of local operators up to precision $\ep\ll 1$:
\bea
\Tr((\rho-\sigma)\mO_p)=\ep O_{\rho,\sigma}(p)
\eea
Then, from the analysis above, we claim that the relative entropy is order $\ep$, which implies that the density matrices are close. One might worry that the sum over infinite terms (the coefficient of $\ep$) can diverge. In this case the relative entropy will diverge which implies that $\rho$ and $\sigma$ have support on unequal subspaces in the Hilbert space. However, in a continuum field theory we believe that all finitely excited energy density matrices are full rank.\footnote{If a density matrix is not full rank it means that the state where it was reduced from can be killed by a local operator with support only on the subsystem, that is the projector to the eigenvector with eigenvalue zero. This violates the ``separating" property of the states of a von Neumann algebra. In the algebraic formulation of quantum field theory, the states are often chosen to be cyclic and separating \cite{Haag:1992hx}.}

 In our case, we are comparing $\psi_{\rm ETH}$ with the reduced state of a thermal density matrix. From~\eqref{eeh}, the 
 one-point functions of local operators in these two states match up to volume suppressed corrections $\ep\sim 1/L$. We thus conclude that the states are close in trace distance up to volume suppressed corrections.

\section{Two dimensional CFTs}\label{sec:2d}

\begin{figure}[b]
\centering
\includegraphics[width=0.6\textwidth]{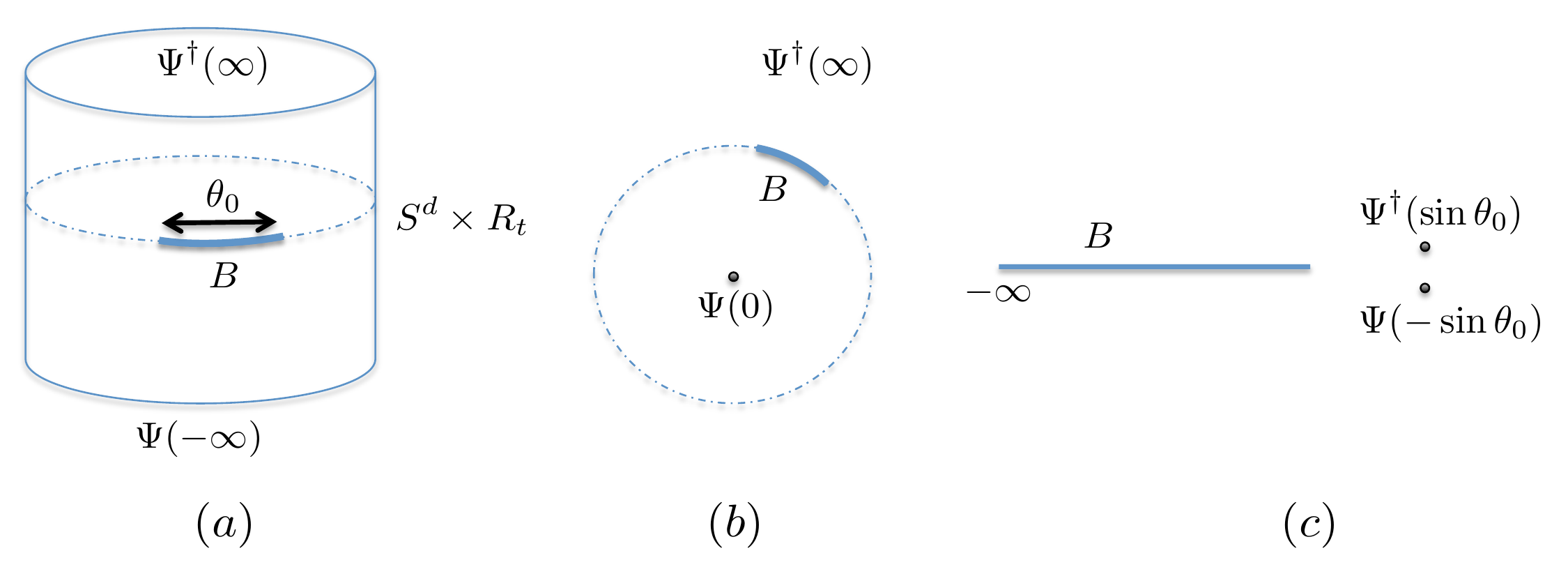}\\
\caption{\small{(a) The cylinder $\mathbb{S}^d\times \mathbb{R}_t$ conformal frame (b) The radial quantization $\mathbb{R}^{d+1}$ conformal frame. (c) The Rindler frame: the conformal frame convenient for the study of the density matrix on subsystem $B$.}}
\label{fig1}
\end{figure}

In this section, we explore the structure of $\psi_{\rm ETH}$~\eqref{uni1}  for a general two-dimensional CFT. 
We show that it is universal across all CFTs of the same central charge. That is to say that the density matrix is comprised of only the polynomials of the stress tensor and the derivative operator, and thus does not depend on any specific structure of a CFT other than the central charge. 
The ETH density matrix ($\psi_{\rm ETH}$) enables us to compute the Renyi and entanglement entropies for primary energy eigenstate.
In next section, we will compare these quantities with those of a generalized Gibbs ensemble.

\subsection{Universal reduced density matrix}  

Consider a two-dimensional CFT on $\mathbb{S}^1 \times \mathbb{R}_t$, where the circle has radius $L$, 
in an energy eigenstate $\ket{\psi}$ of energy $E$.  We take the subsystem $B$ to be an interval of length $2 l$.
We will work with a Euclidean time and it is convenient to use complex coordinates $w = t + i \sig$ with $\sig \in [0,2 \pi L]$. 
 In radial quantization, with $z = e^{w \over L}$, $\ket{\psi}$ and $\bra{\psi}$ are mapped to operators $\Psi (0)$  and $\Psi^\dagger (+\infty)$ of dimension $h = E L$, and 
 $B$ is on the unit circle between $-\theta_0$ and $\theta_0$ with $\theta_0 = {l \over L}$.
The energy density is 
\be 
\ep = {E \ov 2 \pi L} = {h \ov 2 \pi L^2}  \ .
\ee
In the thermodynamic limit we take $L \to \infty$ with $l$ and $\ep$ fixed, and thus $h \propto L^2 \to \infty$. 
We define the thermal length as 
\bea
\label{lmt}
&&\lambda_T=\lb\frac{\la\psi|T_{00}|\psi\ra}{d_T}\rb^{-1/2}=\lb\frac{2\pi h}{c L^2}\rb^{-1/2}, \\
&&d_T=2\la T_{00}T_{00}\ra=\frac{c}{2\pi^2}\nn
\eea
where $T_{00}=\frac{1}{2\pi}(T+\bar{T})$.

 A convenient conformal frame to study the reduced density matrix of $B$ is the Rindler frame in which the subsystem is mapped to the negative half-line see figure \ref{fig1}:
\bea\label{2dmaprind}
&&\omega=\frac{z-q}{q z-1},\quad q=e^{i\theta_0}\nn\\
&&dzd\bar{z}=\Omega(\omega)\bar{\Omega}(\bar{\omega})d\omega d\bar{\omega},\qquad \Omega(\omega)=\frac{(q^2-1)}{(q \omega-1)^2}.
\eea
The operators $\Psi (0)$  and $\Psi^\dagger (+\infty)$ are mapped to $\omega_-=q$ and $\omega_+=q^{-1}$, respectively. 
This is the two dimension version of the map written introduced in \cite{Lashkari:2016vgj};  see Appendix \ref{AppA}.
The key observation of~\cite{Lashkari:2016vgj} is that in the thermodynamical limit, where we take $L \to \infty$ and keep $l$ fixed, $\om_{\pm} \to 1 $ and $\om_- - \om_+ = 2 i \sin\theta_0 \to 0$. The insertions of $\Psi$ and $\Psi^\dagger$ can then be replaced by their OPEs, and the reduced density matrix for region $B$ in the Rindler frame can be written as\footnote{We use tilde to denote density matrices in $\omega$ coordinates: $\tilde{\psi}=U^\dagger\psi U$ where $U$ is the unitary that implements the conformal transformation.}
\be\label{univOPE2d}
\tilde{\psi} =
\Psi(\omega_-,\bar{\omega}_-)\Psi(\omega_+,\bar{\omega}_+)
=\sum_p\sum_{m,n\geq 0}(\omega_--\omega_+)^{h_p+m}(\bar{\omega}_--\bar{\omega}_+)^{\bar{h}_p+n}C^{p,\bar{p},m,n}_{\Psi\Psi} 
\p^m\bar{\p}^n\mO_p 
\ee
where $\mO_p$ is a quasi-primary of dimension $(h_p,\bar{h}_p)$. It should be understood that $\p^m\bar{\p}^n\mO_p $
is inserted at $\om =1$ which we have suppressed. 

The expression~\eqref{univOPE2d} can be further simplified with the following two observations: 
\ben
\item The ratios of the OPE coefficients 
\be 
\frac{C^{p,\bar{p},m,n}_{\Psi,\Psi}}{C^{p,\bar{p},0,0}_{\Psi,\Psi}} 
\ee
is finite (see also Appendix~\ref{AppB}  for explicit expressions). Thus, in the thermodynamic limit the operators with spatial derivatives are $1/L$ suppressed as they are multiplied with extra powers of $(\omega_--\omega_+)^{m}(\bar{\omega}_--\bar{\omega}_+)^{n} \to 0$ for $m,n>0$. 
We can keep only the terms with $m=n=0$. 

\item From~\eqref{ope} the OPE coefficient for quasi-primary $\mO_{p,\bar{p}}$ is given by 
\be 
C^{p,\bar{p}}_{\Psi,\Psi}={L^{(h_p+\bar{h}_p)} \over d_{p,\bar p}} O_{p,\bar p} (E)
\ee
where we have now allowed an arbitrary normalization factor $d_{p,\bar p}$ for two-point function of $\mO_{p,\bar{p}}$. 
We then have 
\be 
(\omega_--\omega_+)^{h_p}(\bar{\omega}_--\bar{\omega}_+)^{\bar{h}_p}C^{p,\bar{p}}_{\Psi\Psi} 
= i^{h_p - \bar h_p} {(2l)^{h_p + \bar h_p} \over d_{p,\bar p}}  O_{p,\bar p} (E) \  
\ee
where we have used that in the thermodynamic limit $2 \sin \theta_0 L = 2 \theta_0 L = 2 l$. 
Local ETH implies that $O_{p, \bar p}$  is, up to corrections suppressed in $L$, the same as the one-point function in the canonical ensemble.  The thermal one-point functions of quasi-primaries which are outside the identity Virasoro block vanish  in the $L\to\infty$ limit as they can be mapped to one-point functions on a complex plane.\footnote{In fact, one can compute the one-point function of primaries on a torus with the modular parameter $\beta/L\ll 1$, and see that the finite-size corrections are exponentially suppressed in volume, see Appendix \ref{AppL}}
This implies that the contribution of any operator outside of the identity Virasoro block vanishes.

\een

We thus conclude that 
 \be \label{vire}
\tilde{\psi} 
\simeq\sum_{(p,\bar{p})\in \text{Viraosoro identitiy block}}
 i^{h_p - \bar h_p} {(2l)^{h_p + \bar h_p} \over d_{p,\bar p}} O_{p,\bar p} (E) 
\mO_p\mO_{\bar{p}}  \ . 
\ee
The Virasoro algebra fixes the dimensions of the operators in the above sum  to  positive integers. We can organize the sum~\eqref{vire} in terms of quasi-primaries 
of dimension $k$ and $\bar{k}$ constructed from the holomorphic (anti-holomorphic) stress tensor and its derivatives. More explicitly,  $\mO_p$ in~\eqref{vire} are given by $\mT_k^{(\alpha)}$'s  which can be schematically written as\footnote{The expression below should be understood as summing over different ways  the derivatives are distributed among $T$'s.}  
\bea
\mT_k^{(\alpha)}= \sum_{k_1+k_2=k} c^{(\al)}_{k_1 k_2} \p^{k_1} T^{k_2} 
\eea
and satisfies the quasi-primary constraint ($L_n$ denote the Virasoro operators)
\bea
L_1\mT^{(\alpha)}_k=0.
\eea
At any positive integer $k$ there are several linearly independent  $\mT_k^{(\alpha)}$ that solve the above quasi-primary constraint, which are labeled by index $\alpha$. 
We show in Appendix~\ref{AppC}, for $k$ even (odd) only one (none) of them survives the thermodynamic limit which is the one with the $T^{k}$ term in it. We take $\alpha=0$ to be the surviving quasi-primary at each level. The same holds for the anti-holomorphic OPE coefficients. Then~\eqref{vire} becomes 
\bea\label{psiuniv2}
\tilde{\psi} 
\simeq \sum_{k,\bar{k}\in \mathbb{N}}
 i^{k - \bar k} {(2l)^{k + \bar k} \over d_{2k} d_{2 \bar k}} O_{k,\bar k} (E) 
\mT^{(0)}_{2k}\bar{\mT}^{(0)}_{2\bar{k}}
\eea
where 
\bea\label{OPEanddp}
&& O_{k,\bar k} (E)  = \la\psi|\mT^{(0)}_{2k}\mT^{(0)}_{2\bar{k}}|\psi\ra, \qquad 
\la \mT^{(0)}_{2k} (z)  \mT^{(0)}_{2k} \ra=\frac{d_{2k}}{|z|^{4k}} \ .
\eea

Operator $\mT^{(0)}_{2k}$ is a polynomial of order $k$ in holomorphic stress tensor $T$ that starts with $T^k \equiv 
(T(T...(T T)))$. The first few $\mT^{(0)}_{2k}$ are computed in Appendix~\ref{AppC}:
\bea
&&\mT^{(0)}_{2}=T,\qquad \mT^{(0)}_{4}=(T T)-\frac{3}{10}\p^2T\nn\\
&&\mT^{(0)}_{6}=(T(TT))+\frac{9(14c+43)}{2(70c+29)}(\p T\p T)-\frac{3(42c+67)}{4(70c+29)}\p^2(T T)-\frac{(22c+41)}{8(70c+29)}\p^4T\nn\\
&&d_2=\frac{c}{2},\qquad d_4=\frac{c(5c+22)}{10}, \qquad d_6=\frac{3c(2c-1)(5c+22)(7c+68)}{4(70c+29)} \ .
\eea

For large $h$, we have
\bea\label{eigenT2k}
&&\la \psi|\mT^{(0)}_{2k}|\psi\ra\simeq\la\psi|T^k|\psi\ra\simeq L^{-2k}\frac{(\mathcal{L}_{-2})^k\la\Psi\Psi\ra}{\la\Psi\Psi\ra}=(h/L^2)^k = \left({c \ov 2 \pi \lam_T^2} \right)^k
\eea
where we have used~\eqref{lmt} and all the other terms in $\mT^{(0)}_{2k}$ are suppressed in $h$:
\bea
\frac{\la\psi|\p^mT|\psi\ra}{\la \psi|T^{1+m/2}|\psi\ra}\sim h^{-m/2}\ll 1.
\eea
We thus find that 
\bea\label{psiETH2d}
\tilde{\psi} 
&\simeq & \sum_{k,\bar{k}\in \mathbb{N}}
 i^{k - \bar k}  \left(\frac{2l}{\sqrt{2\pi}\lambda_T}\right)^{2(k+\bar{k})} \frac{c^{k+\bar{k}}}{d_{2k}d_{2\bar{k}}}
\mT^{(0)}_{2k}\bar{\mT}^{(0)}_{2\bar{k}}. 
\label{red}
\eea
The set of thermodynamically relevant observables are those with non-vanishing expectation value in $|\psi\ra$. From the local ETH we know that this set does not include any operator outside of the Virasoro identity block. 
The translation-invariance of $|\psi\ra$ further implies that among the operators in the identity block only quasi-primaries have a chance of having a non-zero expectation value, because the descendants of quasi-primaries have the derivative operator which are suppressed by $1/L$. The quasi-primaries of dimension $k$ can be organized in the orthonormal basis introduced in appendix \ref{AppC}. Since only $\mT_{2k}$ appear in the universal density matrix $\tilde{\psi}$, they are the only quasi-primaries with non-vanishing expectation value in $|\psi\ra$.


To conclude this subsection we stress that the reduced density matrix~\eqref{red} is universal across all two-dimensional CFTs.

\subsection{Renyi entropies} \label{sec:renyi}



Renyi entropies are invariant under unitary transformations. Hence, we can directly compute them in the Rindler conformal frame. The $n$-th Renyi entropy of a spinless quasi-primary state ($h=\bar{h})$ is given by the Euclidean path-integral over an $n$-sheeted complex plane with $2n$ operators inserted at $q$ and $q^{-1}$ on each sheet.\footnote{Due to the $Z_n$ symmetry of this correlator one can alternatively compute it using a 4-point function with twist operators in a $Z_n$ orbifold theory. This is done in appendix \ref{AppE}.} This manifold is topologically a Riemann sphere, and can be uniformized to one sheet using the map $z=\omega^{1/n}$. Then,
\bea\label{Renyi}
&&\Delta S_n(\psi,l)=\frac{1}{1-n}\log\lb n^{-4n h_\psi} \frac{\la\prod_{j=0}^{n-1}\Psi(z_{j,n},\bar{z}_{j,n})\Psi(z'_{j,n},\bar{z}'_{j,n})\ra}{\la \Psi(z_{0,1},\bar{z}_{0,1})\Psi(z'_{0,1},\bar{z}'_{0,1})\ra^n} \rb\nn\\
&=&\frac{4n h_\psi}{1-n}\log\lb\frac{\sin(\frac{l}{L})}{n\sin(\frac{l}{nL})}\rb+\frac{1}{1-n}\log\lb \frac{\la\prod_{j=0}^{n-1}\Psi_j(z_{j,n},\bar{z}_{j,n})\Psi_j(z'_{j,n},\bar{z}'_{j,n})\ra}{\prod_{j=0}^{n-1}\la \Psi_j(z_{j,n},\bar{z}_{j,n})\Psi_j(z'_{j,n},\bar{z}'_{j,n})\ra}\rb\nn\\
\eea
where $z_{j,n}=e^{i(2\pi j+l/L)/n}$ and $z'_{j,n}=e^{i(2\pi j-l/L)/n}$. 
Using the universal OPE of $\Psi$ in the thermodynamic limit  we find
\bea
&&\Delta S_n(\psi,l)=\frac{(n+1)c}{12\pi n}(2l/\lambda_T)^2+\frac{1}{1-n}\log\Big\la\prod_{j=1}^n\sum_{k_j,\bar{k}_j\in\mathbb{N}}\lb\frac{4c l^2}{2\pi n^2\lambda_T^2}\rb^{k_j+\bar{k}_j} \frac{\mT_{2k_j}(z_{j,n})\mT_{2\bar{k}_j}(\bar{z}_{j,n})}{d_{2k_j}d_{2\bar{k}_j}}\Big\ra\nn .
\eea

\begin{figure}
\centering
\includegraphics[width=0.8\textwidth]{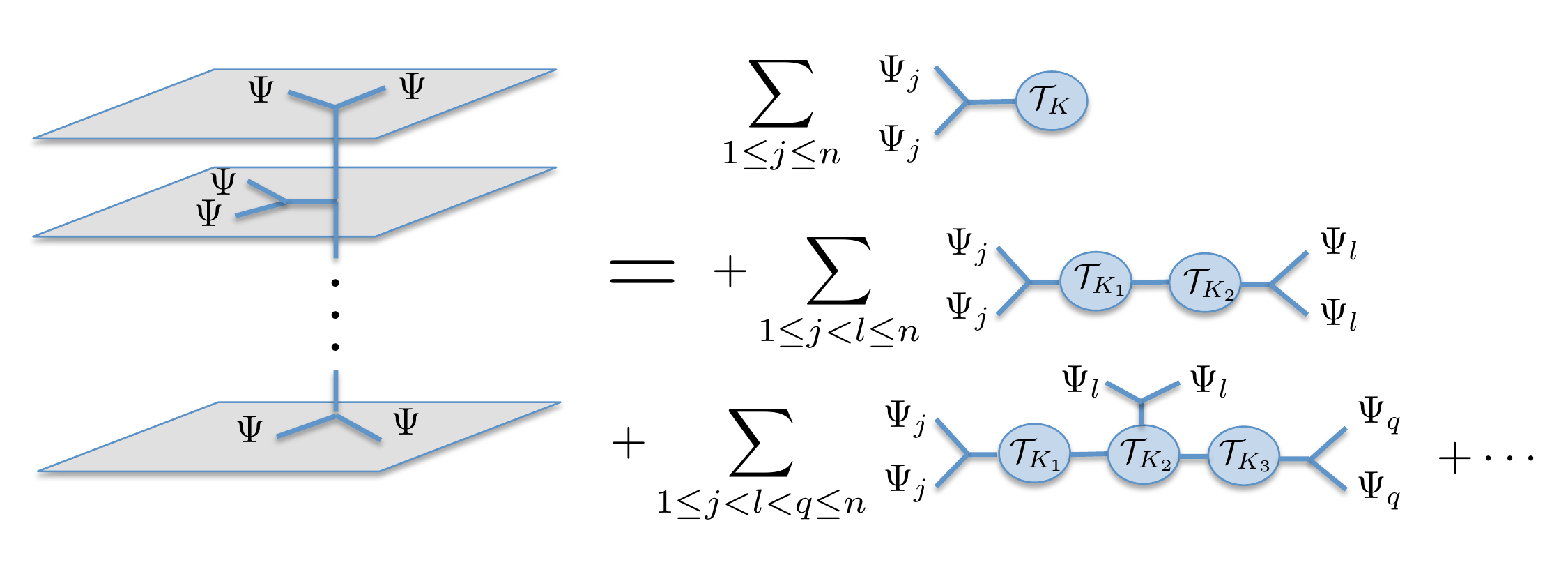}\\
\caption{\small{Renyi entropies correspond to $2n$-point function of the operator that creates the state.}}
\label{fig5}
\end{figure}

Figure \ref{fig5} illustrates the expansion above. The $n$-point functions in the vacuum block are universal. In appendix \ref{AppE} we compute Renyi entropies perturbatively in subsystem size up to order $O\lb(l/\lambda_T)^8\rb$ and find
\bea\label{renyiorder6}
&&\Delta S_n(\psi,l)=\frac{(1+n)c}{12\pi n}(2l/\lambda_T)^2-\frac{(1+n)c}{120\pi^2 n}(2l/\lambda_T)^4\frac{(n^2+11)}{12n^2}\nn\\
&&+\frac{(1+n)c}{630\pi^3 n}(2l/\lambda_T)^6\frac{(4-n^2)(n^2+47)}{144n^4} -\frac{ (1+n)c}{2800 n\pi^4} (2x/\lambda_T)^8 s_8(n,c)+\cdots
\eea
with
\be
s_8(n,c)=\frac{88 (n^2-9) (n^2-4) \left(n^2+119\right)+c \left(-13 n^6+1647 n^4-33927 n^2+58213\right)}{5184 (5 c+22) n^6} \ .
\ee
The authors of \cite{He:2017vyf} computed the Renyi entropies above to the eighth order in the large $c$ limit. The equation (\ref{renyiorder6}) is consistent with their result.

\subsection{Generalized Gibbs ensembles} \label{sec:GGE}

In this section, we explore the relation between the ETH density matrix $\psi_{\rm ETH}$ computed in the last section with that of a Generalized Gibbs Ensemble (GGE). The comparison of observables in these two states 
can be used to study distinguishibility of the corresponding density matrices. Due to the complexity of computing 
the value of observables in a GGE, our comparison is, so far, incomplete. We hope this discussion can set the stage for future investigations of 
the properties of GGE.

Two-dimensional CFTs have an infinite number of conserved charges, which are the KdV hierarchy of charges
$\{Q_{2k-1}, \bar Q_{2k-1}, \; k=1,2, \cdots\}$  
 constructed from the polynomials of stress tensor~\cite{bazhanov1996integrable,francesco2012conformal} 
\bea
&&Q_{2k-1}=\frac{1}{2\pi i}\oint d\omega J_{2k}(\omega)\ ,\nn\\
&&[Q_{2k-1},Q_{2l-1}]=0\ ,
\eea
with the first few local currents given by 
\bea
&&J_2=T, \quad J_4=(T T),\quad J_6=(T(T T))-\frac{c+2}{12}(\p T\p T) .
\eea
On a cylinder of circumference $2\pi L$ the first two charges are
\bea\label{ConservedCharge}
&&Q_1=\frac{1}{L} \lb L_0-\frac{c}{24}\rb\nn\\
&&Q_3=\frac{1}{L^3} \lb 2\sum_{n=1}^\infty L_{-n}L_n+L_0^2-\frac{c+2}{12}L_0+\frac{c(5c+22)}{2880}\rb\ .
\eea
A  Virasoro primary $\ket{\psi}$ is a simultaneous eigenstate of $\{Q_{2k-1}, \bar Q_{2k-1}, \; k=1,2, \cdots\}$, with 
all the eigenvalues $\{q_{2k-1}, \bar q_{2k-1}, \; k=1,2, \cdots\}$ fixed in terms of only the conformal dimension $h_\psi  = E L $. For example, the charges associated to $Q_1$ and $Q_3$ are
\be\label{KdVprimary}
q_1 = L^{-1} \lb h_\psi-\frac{c}{24}\rb , \quad q_3 = L^{-3} \lb h_\psi^2-\frac{c+2}{12}h_\psi+\frac{c(5c+22)}{2880}\rb \ .
\ee

In what follows we assume that the hypothesis of local ETH ~\eqref{ETH} holds for any sufficiently excited Virasoro primary $\Psi$. 
As we discussed in Sec.~\ref{sec:nee}, we expect that $\psi_{\rm ETH}$ for the eigenstate $\ket{\psi}$ prepared from $\Psi$ to be close in trace distance to 
the reduced density matrix of the GGE 
\bea\label{GGEmatrix}
\rho_{GGE}=Z^{-1}\: \exp\lb -\sum_{k=1}^\infty\lb \mu_{2k-1}Q_{2k-1}+\bar{\mu}_{2k-1}\bar{Q}_{2k-1}\rb\rb,
\eea
where the chemical potentials $\{\mu_{2k-1}, \bar \mu_{2k-1}\}$ are chosen to match the set of charges $\{q_{2k-1}, \bar q_{2k-1}\}$ of $\ket{\psi}$. 
If correct, \eqref{GGEmatrix} would  provide a non-trivial consistency check of the local ETH hypothesis. 
In the thermodynamic limit the KdV charges of a Virasoro primary are easy to compute: 
\be\label{1pteigens}
{q_1 \ov L} =  2 \pi \ep ,\quad  {q_3 \ov L}= (2 \pi \ep)^2 , \quad \cdots , \quad  {q_{2k-1} \ov L}= (2 \pi \ep)^k , \quad \cdots \ 
\ee
where $\ep$ is the energy density.

To proceed further, we assume that the central charge $c$ is large. In the $c \to \infty$ limit, all $\mu_{2k-1}$ except $\mu_1 = \beta$ vanish;  
thus we recover the standard Gibbs ensemble \cite{deBoer:2016bov,Basu:2017kzo}. To see this, note that in the large $c$ limit (see the next subsection for a derivation)
\be \label{thermalcor}
{1 \ov Z} \Tr \left(J_{2k} e^{-\beta H} \right) =   \left({\pi^2 c \ov 6 \beta^2 } \right)^k. 
\ee
The two-dimensional thermal energy density is 
\bea
\ep(\beta)=\frac{c \pi}{6\beta^2}.
\eea
Matching this with the energy density in the eigenstate, using (\ref{uj}) and the definition of $d_T$ in (\ref{lmt}), we find
\be\label{changepara}
\lambda^2=\frac{3\beta^2}{\pi^3}\ .
\ee
In the next subsection, we use this change of parameters to compare the one-point functions in the energy eigenstate in (\ref{eigenT2k}) with those of the thermal state in (\ref{thermalcor}) in the $c\rightarrow \infty$ limit and they match exactly. 

The reduction of the conventional Gibbs ensemble $e^{-\beta H}/Z$ is only matching the ETH density matrix in the infinite central charge limit. The necessity to modify it when the central charge is finite is suggested by the non-zero values of KdV charges \eqref{1pteigens}. Historically, first indication that the excited primary state is locally different from thermal state came from the comparison of entanglement and thermal entropies in \cite{He:2017vyf}, although it should be noted that such a discrepancy by itself does not immediately preclude the corresponding density matrices to be trace-distance close \cite{Lashkari:2016vgj,Dymarsky:2016aqv}. A direct comparison of local observables unambiguously showing that ETH density matrix can not match the canonical one was soon performed in \cite{Basu:2017kzo}, with more analysis probing finite $1/c$ corrections in an attempt to match ETH density matrix with the GGE one following in \cite{He:2017txy}. 
In this paper we further investigate this question.  The main unresolved challenge here is to compute the expectation value of KdV currents in the GGE at finite $c$. Despite the fact that for larger $k$ corresponding $\mu_{2k-1}$ are suppressed by the increasingly negative powers of $c$,  we find a strong indication that one cannot perform a perturbative analysis by truncating~\eqref{GGEmatrix} to finite number of $\mu$'s even for the next to the leading order in the $1/c$ expansion (see Appendix \ref{nonperturb}).  Hence to complete the check, one needs a truly non-perturbative expression for \eqref{GGEmatrix} both in terms of powers and numbers of included $\mu$'s. We leave this task for a future investigation.

In the limit $h_a\gg c\gg 1$, the universal density matrix in (\ref{psiETH2d}) simplifies and exponentiates (see Appendix~\ref{AppC})
\bea
&& \tilde \psi =e^{(D_a T+\bar{D}_a \bar{T})} ,\qquad a^2=\frac{(2\pi L)^2}{12\beta^2} \\
&&D_a=a^2-\frac{a^4}{10\times 2!}\p^2+\frac{11a^6}{70\times 4!}\p^4-\frac{9a^8}{140\times 6!}\p^6-\frac{34a^{10}}{1925\times 8!} \p^8 +\cdots \ .
\eea
This is because at large $c$
\be
{1 \ov d_{2k}} \simeq {2^k \ov k! c^k},
\ee
and we have used the change of parameters in (\ref{changepara}).
Note that in order to properly define the operator $\tilde{\psi}$ one has to smooth out the exponent on a circle of radius $\ep$ around $z=0$ where the operator is inserted:
\bea
\la \tilde{\psi} \cdots\ra=\la e^{\oint_{r=\ep} D_a T+\bar{D}_a\bar{T}}\cdots \ra
\eea

\subsection{Matching with thermal density matrix in the infinite $c$ limit} 
\label{sec:infinitec}
In two dimensions, the thermal cylinder is conformally flat, therefore the expectation value of any operator that is outside of the Virasoro identity block vanishes in the Gibbs state. The translation-invariance further restricts the set of observables with non-vanishing thermal one-point function to the quasi-primaries. 
Below, we show that at large $c$ the thermal expectation value of $\mT_{2k}$ scales as $c^k$, whereas the expectation value of other quasi-primaries of the same conformal dimension scale with lower powers of $c$.  

The current $J_{2k}$ is a polynomial of order $k$ in stress tensor, where the normal-ordered operator  $(T^k)=(T(T\cdots (T T)))$ is defined by isolating the distance independent term in the OPE:
\bea
&&(A B)(\omega)=\lim_{\omega_1\to \omega}\lb A(\omega_1)B(\omega)-\text{divergent terms}\rb\nn\\
&&=\frac{1}{2\pi i}\oint \frac{d\omega_1}{(\omega-\omega_1)} A(\omega_1)B(\omega),
\eea
where in the second line the normal ordering is imposed by a Cauchy integral. In a thermal state 
\bea\label{cauchy}
&&tr(\rho_\beta (T T)(\omega))=\frac{1}{2\pi i}\oint_{\omega} \frac{d\omega_1}{\omega_1-\omega}\:tr(\rho_\beta T(\omega_1)T(\omega))\nn\\
&&tr(\rho_\beta(T (T T))(\omega))= \frac{1}{2\pi i}\oint_{\omega}\oint_{\omega} \frac{d\omega_1 d\omega_2}{(\omega-\omega_1)(\omega-\omega_2)}\:tr(\rho_\beta T(\omega_1)T(\omega_2) T(\omega)).
\eea
At large central charge, the multi-point thermal correlators are dominated by the disconnected piece:
\bea
tr(\rho_\beta T(x_1)\cdots T(x_k))=tr(\rho_\beta T(x))^k \lb 1+O(1/c)\rb= \lb \frac{\pi^2 c}{6\beta^2}\rb^k \lb 1+O(1/c)\rb.
\eea
Plugging this in the right hand side of (\ref{cauchy}), and performing the Cauchy integral we obtain
\bea
tr(\rho_\beta (T^k))= \lb \frac{\pi^2 c}{6\beta^2}\rb^k \lb 1+O(1/c)\rb.
\eea

Now, consider a quasi-primary that is not $\mT_{k}$. The first non-trivial such quasi-primary appears at dimension six: 
\bea
A=(\p T\p T)-\frac{2}{9}\p^2 (T T)+\frac{1}{42}\p^4 T.
\eea
The normal-ordering is imposed by 
\bea
tr(\rho_\beta A(\omega))=tr(\rho_\beta (\p T \p T)(\omega))=\frac{1}{2\pi i}\oint_{\omega} \frac{d\omega_1}{\omega_1-\omega}\:tr(\rho_\beta \p T(\omega_1)\p T(\omega))=O(c),
\eea
where we have used the fact that the thermal state is translation-invariant in space and time; hence, the disconnected piece of the expectation value on the right hand side is zero. The same conclusion applies to all other quasi-primary operators in the Virasoro identity block that are not $\mT_{2k}$, as they also can be considered as multi-trace operators with at least one factor containing derivatives.
If we redefine the stress tensor in the large $c$ limit according to $\tilde{T}=T/c$, the expectation value of $\mT_{2k}$ become order one, while the expectation value of any other quasi-primary in the Virasoro identity block is suppressed by negative powers of $c$. Thus, the only operators with non-vanishing expectation values in this limit are $\mT_{2k}$.

The quasi-primary operators $\mT_{2k}$ and the KdV charges $J_{2k}$ are both polynomials of order $k$ in $T$ and start with $(T^k)$.
The derivative terms are different, however, as we just discussed the derivative terms are suppressed in large central charge. Therefore,
\bea
&&tr(\rho_\beta \mT_{2k})=tr(\rho_\beta J_{2k})=\lb \frac{\pi^2 c}{6\beta^2}\rb^{k}\lb 1+O(1/c)\rb
\eea
This is the same answer as the one-point functions in the eigen-state (\ref{eigenT2k}) after we replace $\beta^2= \pi^3\lambda_T^2/3$.
Since there are no other thermodynamically-relevant observables we have found that all the one-point functions of the eigenestate matches of those of the Gibbs state in the large central charge limit. Thus, in the large $c$ limit we have proved that the universal density matrix of a Virasoro primary eigenstate is indistinguishable from  
that of the Gibbs state.

It is interesting to compare the Renyi entropies in the thermal state with the eigenstate in the large $c$ limit. We can take a large $c$ limit in the low temperature expansion in (\ref{renyiorder6}) the perturbation theory of small $x/\lambda_T$:
\bea\label{sixthent2}
&&\Delta S_n(\psi,x)=\frac{(1+n)c}{12n\pi}(2x/\lambda_T)^2-\frac{(1+n)c}{120 n\pi^2}\frac{(n^2+11)}{12n^2}(2x/\lambda_T)^4\nn\\
&&+\frac{(1+n)c}{630n\pi^3}\frac{(4-n^2)(n^2+47)}{144n^4}(2x/\lambda_T)^6
-\frac{ (1+n)c}{2800 n\pi^4} (2x/\lambda_T)^8 s_8+\cdots
\eea
with
\be
s_8=\frac{-13 n^6+1647 n^4-33927 n^2+58213}{25920 n^6} \ .
\ee
It is clear that the Renyi entropies for $n>1$ do not acquire thermal values given by 
\bea
&&\Delta S_n(\beta,x)=\frac{(n+1)c}{6}\log\lb\frac{\beta}{2\pi x} \log\lb \frac{2\pi x}{\beta}\rb\rb\nn\\
&&=\frac{(1+n)c}{12n\pi}(2x/\lambda_T)^2-\frac{(1+n)c}{120 n\pi^2}(2x/\lambda_T)^4+\frac{(1+n)c}{630n\pi^3}(2x/\lambda_T)^6+\cdots 
\eea
where in the second line we have used the change of parameters in (\ref{changepara}). This is in contrast with the entanglement entropies of the states that match to the eighth order that we have computed.  

In the large $c$ limit, one can in fact compute the dominant $c$ piece of the entanglement entropy of the eigenstate non-perturbatively for finite values of $l/\lambda_T$. In section~\ref{sec:renyi}, we computed the Renyi entropies directly by constructing the partition function that represents $tr(\rho^2)$ and uniformizing it. An alternative method to compute the Renyi entropy of the eigenstate is computing the four-point function of twist operators with $\Psi^n$ in an orbifold theory; see (\ref{sigmasigma}) of Appenix~\ref{AppE}. The assumption of local ETH tells us that only the Virasoro identity block contributes to the correlator
\bea
G_4(z,\bar{z})=\la \Psi^n(\infty) \sigma_n(z,\bar{z})\sigma_n(1)\Psi^n(0)\ra.
\eea 
where $z=e^{i x/L}$. 
The leading $c$ piece of the contribution of the Virasoro identity block to the four point function above in the large $c$ limit was found by solving the monodromy equation for $n$ near $n=1$ in~\cite{Asplund:2014coa}:
\bea
&&\log G_4(z,\bar{z})\simeq\frac{c(1-n)}{6}\log\lb \frac{z^{(1-\alpha_\psi)/2}\bar{z}^{(1-\bar{\alpha}_\psi)/2}(1-z^{\alpha_\psi})(1-\bar{z}^{\bar{\alpha}_\psi})}{\alpha_\psi\bar{\alpha}_\psi}\rb+O((n-1)^2)\nn\\
&&\alpha_\psi\equiv i\sqrt{\frac{h_\psi}{24}}.
\eea

 The entanglement entropy computed this way from the identity block in the large $c$ limit matches the entanglement entropy in the Gibbs state for any $l/\beta$. Note that here we are working in the limit where $h_\psi\gg c\gg 1$, which in the language of \cite{Asplund:2014coa} translates to $h_\psi=\alpha c$, $c\gg 1$ and $\alpha\gg 1$. In our approach the assumption of local ETH guarantees that only the Virasoro identity block dominates. However, the authors of \cite{Asplund:2014coa}  assumed a sparse spectrum of low-dimension operators to truncate to the identity block.

\section{Higher dimensional CFTs} \label{sec:hd}

In this section, we first discuss the general structure of the ETH density matrix in higher dimensions, and then 
compute the entanglement entropy to the leading nontrivial order in $l/\lam_T$ expansion. We compare the result to the holographic entanglement entropy computed using the Ryu-Takayanagi formula at this order and find agreement. The intuition is that even though our CFT computation does not assume large $N$ or strong coupling, at this order the answer is universal because it depends only on $d_T$ that is the normalization of the two-point function of stress tensor. 
To match the entanglement entropies we have to set the coefficient $d_T$ to be (\ref{dTholo}), as is required in a holographic CFT.
This provides a consistency check of the local ETH. 

\subsection{ETH density matrix} 

We observed that in two dimensions assuming local ETH implies that only the polynomials of stress tensor propagate in the thermodynamic limit of OPE.
Here, we consider density matrices in primary energy eigenstates of higher-dimensional CFTs satisfying local ETH.
A generalization of the map introduced in (\ref{2dmaprind}) (see appendix \ref{AppA}) maps the radial quantization frame to the Rindler frame.  In Rindler coordinates, the subsystem $B$ is mapped to the negative half-space $X_1<0$, and the operators that create and annihilate the state are, respectively, at $X_\mu^-$ and $X_\mu^+$. Since $X_{i>2}^{\pm}=0$ we can use the two-dimensional complex coordinates to describe their location:  
$X_0^-+i X_1^-=e^{-i\theta_0}=1/q$ and $X^+_0+i X_1^+=e^{i\theta_0}=q$.\footnote{Note that compared to the two-dimensional map the location of $\omega_-$ and $\omega_+$ are swapped.} The distance between the two operators in these coordinates is $2\sin\theta_0\simeq 2l/L$. The operator product expansion in the thermodynamic limit $l/L\to 0$ becomes
\bea\label{OPEhigher}
&&\frac{\Psi(X^+_\mu)\Psi(X^-_\mu)}{\la\Psi(X^+_\mu)\Psi(X_-)\ra}\simeq \sum_{p}^\infty C_{\psi\psi}^{p,\hat{n}}\:|\vec{n}|^{h_p}\mO_{p}^{\hat{n}}(X^-_\mu)=\sum_{p}^\infty f_p^{\hat{n}} (l/\lambda_T)^{h_p} \mO_{p}^{\hat{n}}(X^-_\mu)\nn\\
\eea
where $X^+_\mu=2\sin\theta_0 \hat{n}_\mu+X^-_\mu$, $\hat{n}$ is the unit vector in the $X_0$ directions, and we have dropped the descendant fields because their contribution is $1/L$ suppressed. The operator $\mO_p^{\hat{n}}$ is  a primary with spin with its indices contracted with $\hat{n}$ according to
\bea\label{fp}
&&\mO_{p}^{\hat{n}}=\lb \hat{n}^{\mu_1}\hat{n}^{\mu_2}\cdots -\text{traces} \rb (\mO_p)_{\mu_1\mu_2\cdots}\nn\\
&&C^{p,\hat{n}}_{\psi\psi}=\frac{\la \Psi(\infty)\mO_{p}^{\hat{n}}(1)\Psi(0)\ra}{\la \mO_p^{\hat{n}}(1)\mO_p^{\hat{n}}(0)\ra \la \Psi(\infty)\Psi(0)\ra},
\eea
and finally $f_p$ is defined by
\bea\label{fp2}
&&f_{p}^{\hat{n}}=(2\lambda_T/L)^{h_p}C^p_{\psi\psi}=\frac{\la\psi|\mO_{p}^{\hat{n}}|\psi\ra}{d_{p,\hat{n}}}(2\lambda_T)^{h_p} \nn\\
&&d_{p,\hat{n}}=\la \mO_{p}^{\hat{n}}(1)\mO_{p}^{\hat{n}}(0)\ra.
\eea
It is customary to define a coefficient $d_p$ that is independent of $\hat{n}$ in the following way:
\bea
&&\la (\mO_p)_{\mu_1\cdots \mu_m}(x^\mu)(\mO_{p'})_{\nu_1\cdots \nu_m}(0)\ra=d_{p} \delta_{p p'} |x|^{-2h_p}I_{\mu_1\cdots \mu_m,\nu_1\cdots\nu_m},
\eea
where the tensor $I_{\mu_1\cdots \mu_m,\nu_1\cdots \nu_m}$ is fixed by conformal symmetry \cite{Osborn:1993cr,Rychkov:2016iqz}. Every CFT has a stress tensor that is a primary of dimensions $d+1$. The energy density in primary state $|\psi_a\ra$ is
\bea
\ep=\frac{E}{L^d\omega_d}=\frac{h_a}{L^{d+1}\omega_d},
\eea
where $\omega_d$ is the volume of the unit sphere $\mathbb{S}^d$. 
As an example, consider the term in the OPE expansion (\ref{OPEhigher}) that corresponds to stress tensor 
\bea
&&\frac{\ep}{d_{T,\tau}} (2l)^{(d+1)} \lb \hat{n}^{\mu}\hat{n}^{\nu}-\delta^{\mu\nu}/(d+1) \rb T_{\mu\nu}\nn\\
&&=\frac{d+1}{d}\lb\frac{2l}{\lambda_T}\rb^{(d+1)}\lb \hat{n}^{\mu}\hat{n}^{\nu}-\delta^{\mu\nu}/(d+1) \rb T_{\mu\nu},\nn\\
&&\lambda_T=\lb \frac{\ep}{d_T}\rb^{-1/(d+1)}
\eea
where $\lambda_T$ is the length associated with the energy density, and $d_T$ is the central charge defined by the two-point function of stress tensor:
\bea\label{dT}
&&\la T_{\mu\nu}(u)T_{\alpha\beta}(v)\ra=\frac{d_T}{|u-v|^{2(d+1)}}S_{\mu\nu,\alpha\beta}(u-v),\nn\\
&&S_{\mu\nu,\alpha\beta}(u)=\frac{1}{2}\lb I_{\mu\alpha}(u)I_{\nu\beta}(u)+(\mu\leftrightarrow\nu)\rb-\frac{1}{d+1}\delta_{\mu\nu}\delta_{\alpha\beta}\nn\\
&&I_{\mu\nu}(u)=\delta_{\mu\nu}-\frac{2u_\mu u_\nu}{|u|^2}.
\eea

To obtain the density matrix in the thermodynamic limit we have to study the OPE in (\ref{OPEhigher}) in more detail. From the equivalence of the microcanonical ensemble and the thermal ensemble we expect the coefficient
\bea\label{fpthermal}
f_p\simeq \frac{(2\lambda_T)^{h_p}}{d_p}tr(\rho_T\mO_p)
\eea
to have the interpretation of a thermal one-point function up to volume suppressed corrections, where the thermal state is chosen to have the same energy density as the eigenstate $|\psi\ra$. In two dimensions, we saw that thermal one-point functions vanish which let to a truncation of the OPE to only the Virasoro identity block.
However, in higher than two dimensions thermal one-point functions do not vanish, and $f_p$ are, potentially, non-zero. 

One way to obtain universality in higher dimensions is by restricting the class of higher dimensional theories we study; for instance the holographic theories.
In holographic CFTs the thermal one-point function of conformal primaries are $1/N$ suppressed except for operators constructed from the stress tensor. Tn large $N$ CFTs resemble two-dimensional CFTs in the sense that they have multi-trace operators $T^m$ in their spectrum that are primaries of conformal dimension $m(d+1)$, up to $1/N$ corrections. In holographic theories the thermal correlator is essentially classical, that is to say the thermal variance of the operator $T$ is $1/N$ suppressed:
\bea
tr(T^2 \rho_T)-tr(T\rho_T)^2=O(1/N)
\eea
Therefore, from local ETH and the equivalence of ensembles one expects $C^{T^m}_{\psi\psi}\sim h^m$ which implies that they survive the thermodynamic limit and contribute to $\mathcal{A}_{therm}$. In holographic theories, $T^m$ are in $\mathcal{A}_{therm}$ and one needs to include them in the sum in the definition of the ``universal" density matrix.\footnote{At finite central charge the only primaries one can construct from $T$ are large spin operators of type $(T T)_{n,l}\equiv T\p_{\mu_1}\cdots \p_{\mu_l}(\p^2)^nT$. 
In fact for large $l$ there are operators of this type for all $m\in\mathbb{N}$: $(T^m)_{(n_1,l_1)\cdots (n_m,l_m)}=((T T)_{n_1,l_1}T_{n_2,l_2})\cdots  T_{n_,l_m})$. However, every derivative suffers a $1/L$ suppression and hence one expects their OPE coefficients to scale, at best, as $h^{m}$ rather than $h^{m+(2n+l)/(d+1)}$ that is required to survive the thermodynamic limit. An explicit calculation of the OPE coefficients $C^{[TT]_{n,l}}_{\psi\psi}$ confirms this expectation \cite{Fitzpatrick:2015qma}. This calculation is done assuming that the spin is largest parameter. However, for our case of interest we want the conformal dimension of the operator to be much larger than its spin which is much larger than one. It is plausible that in our limit of interest these operators survive the thermodynamic limit and contribute to $\mathcal{A}_{therm}$. We thank Liam Fitzpatrick and Sasha Zhiboedov for pointing this out to us. } 


\subsection{Entanglement entropy from ETH density matrix} 

As opposed as two-dimensional case, the ETH density matrix in (\ref{OPEhigher}) is not universal. That is to say that at finite central charge we only know one operator in the set of thermodynamically relevant operators $\mathcal{A}_{therm}$. If we try to repeat our low temperature analysis of the ETH density matrix in $d>2$ we need to make further assumptions about the spectrum of the theory.

Let us assume that there are no relevant primary operators in the set $\mathcal{A}_{therm}$. In other words, we are assuming that $f_p=0$ for all  operators $\mO_p$ in(\ref{OPEhigher}) with $h_p<d$. Then, to the first non-trivial order the ETH density matrix is
\bea\label{lowtempd}
&&tr(\tilde{\psi}\cdots )\simeq \left\la \lb1+\lb\frac{d+1}{d}\rb\lb\frac{2l}{\lambda_T}\rb^{d+1} \hat{n}^\mu \hat{n}^\nu T_{\mu\nu}+\cdots\rb \cdots \right\ra \ .
\eea
In a CFT the operator $T_{\mu}^{\:\:\mu}=0$ in flat space. 
Now, we  can compute the entanglement entropy of the ETH density matrix at this order and compare it with the reduced density matrix of the Gibbs state.
%
Renyi entropies are unitarily invariant, and it is more convenient to compute the entanglement entropy in Rindler coordinates. The vacuum-subtracted Renyi entropy in primary state $|\psi\ra$ is given by
\bea
&&\Delta S_n(\psi,l)=\frac{1}{1-n}\log\frac{\la\prod_{j=1}^n\psi_j\psi_j\ra_{(2\pi n)}}{\la\psi\psi\ra_{(2\pi)}}
\eea
where the subscript $(2\pi n)$ refers to the angle around the boundary of $B$: $X_0=X_1=0$ in Rindler space. We denote the generator of rotation around this hypersurface as $\p_\tau$:
\bea
X_0+i X_1=\omega,\: \omega/\bar{\omega}=e^{2i\tau}.
\eea
We are interested in entanglement entropy which is found from the $n\to 1$ limit of
\bea\label{changeEntropy}
&&\Delta S(\psi,l)=\delta^{(1)}S+\delta^{(2)}S\nn\\
&&\delta^{(1)}S=-\p_n\left[ n\log\frac{\la\psi\psi\ra_{(2\pi n)}}{\la\psi\psi\ra_{(2\pi)}}\right]_{n\to 1}\nn\\
&&\delta^{(2)}S=-\p_n\log \left[\frac{\la\prod_{j=1}^n\psi_j\psi_j\ra_{(2\pi n)}}{\la \psi\psi\ra^n_{(2\pi n)}} \right]_{n\to 1}.
\eea
Our calculation closely follows the method used in \cite{Faulkner:2014jva}, and uses the Hamiltonian language:
\bea
\la\psi\psi\ra_{(2\pi n)}=tr\lb e^{-2\pi n H}\mathcal{P}(\psi \psi)\rb
\eea
where $\mathcal{P}$ is the path-ordering operator in the Euclidean space.
The first term in (\ref{changeEntropy}) is the change in the expectation value of the vacuum modular operator $H$:
\bea
&&\delta^{(1)}S=-\frac{\p_ntr(e^{-2\pi n H} \mathcal{P}(\psi\psi))\Big|_{n\to 1}}{\la\psi\psi\ra_{(2\pi)}}=\frac{2\pi\la H\psi\psi\ra_{(2\pi)}}{\la\psi\psi\ra_{(2\pi)}}=\frac{2\pi \omega_{d-1}\ep \: l^{d+1}}{d(d+2)}=\frac{2\pi \omega_{d-1}d_T}{d(d+2)}\lb\frac{l}{\lambda_T}\rb^{d+1}\nn\\
&&\omega_{d-1}=\frac{2\pi^{d/2}}{\Gamma(d/2)}.
\eea
This is the so-called first law of entanglement entropy; for small variation of density matrix $\delta S=2\pi \delta H$, where $H$ is the generator of Euclidean rotation in the $\tau$ direction. The second term in (\ref{changeEntropy}) is the relative entropy of the eigenstate with respect to the vacuum reduced to the subsystem $B$: $S(\psi\|\sigma).$
The task is to compute the relative entropy above perturbatively in powers of $l/\lambda_T$.

\begin{figure}
\centering
\includegraphics[width=0.8\textwidth]{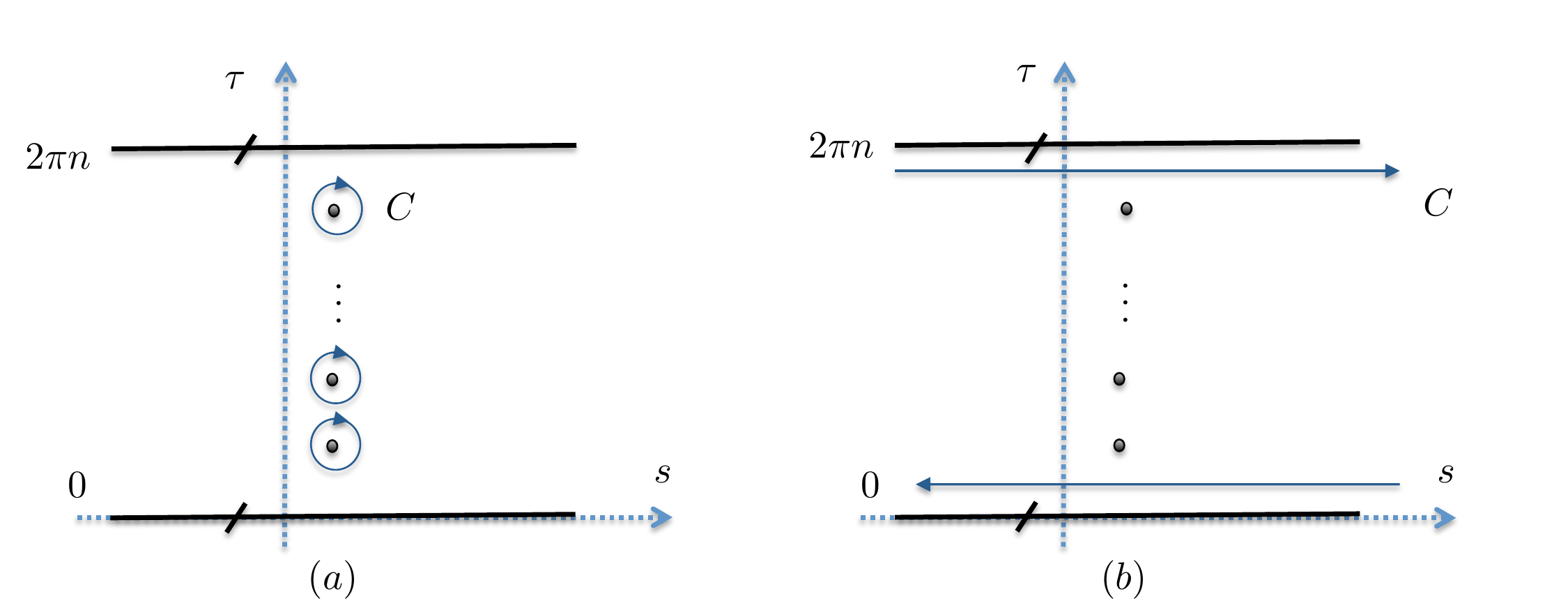}\\
\caption{\small{ (a) The path-integral over complexified $\tau$ picks up $n$ poles at $\tau=2\pi j$. (b) The contour $C$ is deformed to run over $(-\infty+i(2\pi n-\ep),\infty+i(2\pi n-\ep))$ and $(\infty+i\ep,-\infty+i\ep)$ .}}
\label{fig7}
\end{figure}

Since $\Psi$'s approach each other pairwise in Rindler space, one can use the flat space OPE. At the next-to-leading order the entanglement entropy is
\bea
&&\delta^{(2)}S=\frac{(d+1)^2}{d^2}\lb\frac{2l}{\lambda_T} \rb^{2(d+1)}\p_n\left[\frac{n}{2}\sum_{j=1}^{n-1} G_n^{00}(2\pi j)\right]_{n\to 1}\nn\\
&&G_n^{\mu\nu}(2\pi j)=\la T_{\mu\mu}(0)T_{\nu\nu}(2\pi j)\ra_{(2\pi n)}.
\eea
where the index $0$ signifies the $X^0$ in Rindler coordinates.
We follow the method advocated in \cite{Faulkner:2014jva} to analytically continue the expression above in $n$:
\bea
&&A^{\mu\nu}(n)=\sum_{j=1}^{n-1}G_n^{\mu\nu}(2\pi j)=\int_C \frac{ds}{2\pi i} \frac{G_n^{\mu\nu}(-is)}{e^s-1}\nn
\eea
where $s$ is the complexified $\tau$ angle. The contour $C$ is deformed to run over $(-\infty+i(2\pi n-\ep),\infty+i(2\pi n-\ep))$ and $(\infty+i\ep,-\infty+i\ep)$; see figure (\ref{fig7}):
\bea
&&A^{\mu\nu}(n)=\int_{-\infty}^\infty \frac{ds}{2\pi i}\lb \frac{G_n^{\mu\nu}(-is+\ep)}{e^{s+i\ep}-1}-\frac{G_n^{\mu\nu}(-is+2\pi n-\ep)}{e^{s+2\pi in-i\ep}-1}\rb
\eea

The analytic continuation is the choice to set $e^{2\pi i n}=1$ in the denominator.
\bea
\p_nG_n^{\mu\nu}(-is+\ep)\big|_{n\to 1}=\p_ntr\left[e^{-2\pi n H}T_{\mu\mu}(0)T_{\nu\nu}(s+i\ep)\right]_{n\to 1}=-2\pi tr\left[e^{-2\pi H}H T_{\mu\mu}(0)T_{\nu\nu}(s+i\ep)\right]\nn
\eea
and
\bea
\p_nA(n)^{\mu\nu}\big|_{n\to 1}=i\int_{-\infty}^\infty ds\left[\frac{tr(e^{-2\pi H}H T_{\mu\mu}(0)T_{\nu\nu}(s+i\ep))}{e^{s+i\ep}-1}-\frac{tr(e^{-2\pi H}H T_{\mu\mu}(s-i\ep)T_{\nu\nu}(0))}{e^{s-i\ep}-1}\right]\nn
\eea
The second term can be further simplified using the commutator $[H,T_{\mu\mu}(s)]=-i \frac{dT_{\mu\mu}}{ds}$ and the KMS condition
\bea
&&tr(e^{-2\pi H}H T_{\mu\mu}(s-i\ep)T_{\nu\nu}(0))=tr(e^{-2\pi H}(T_{\mu\mu}(s-i\ep)H-[H,T_{\mu\mu}(s-i\ep)]) T_{\nu\nu}(0))\nn\\
&&\nn\\
&&=tr(e^{-2\pi H}HT_{\mu\mu}(0)T_{\nu\nu}(s+2\pi i-i\ep))+i \frac{d}{ds} tr(e^{-2\pi H}T_{\mu\mu}(s-i\ep) T_{\nu\nu}(0))\nn
\eea
Putting this back in $A(n)$ gives
\bea
\p_nA^{\mu\nu}(n)\big|_{n\to 1}&&=i\int_{-\infty}^\infty ds\lb \frac{G_1^{\mu\nu}(-is+\ep)}{e^{s+i\ep}-1}-\frac{G_1^{\mu\nu}(-is-\ep)}{e^{s-i\ep}-1}\rb\nn\\
&&+\int_{-\infty}^\infty \frac{ds}{e^{s-i\ep}-1}\frac{d}{ds}tr(e^{-2\pi H}T_{\mu\mu}(s-i\ep) T_{\nu\nu}(0))
\eea
The term in the first line vanishes since there are no poles in the region encircled by the contour integration. Using integration by parts we can write the second term as
\bea
\p_nA^{\mu\nu}(n)\big|_{n\to 1}=-\int_{-\infty}^\infty \frac{ds}{4\sinh^2((s-i\ep)/2)}\la T_{\mu\mu}(X_s)T_{\nu\nu}(X_0)\ra
\eea
where $X_0=(1,-i s/2,\cdots)$ and $X_s=(1,i s/2,0,\cdots)$ in Rindler coordinates. 
Therefore,
\bea
\p_n A^{\mu\nu}(n)\big|_{n\to 1}=(-1)^{d+1}\int_{-\infty}^\infty \frac{ds\:d_T}{(2\sinh(\tilde{s}/2))^{2(d+2)}}S_{\mu\mu,\nu\nu}\nn
\eea
where $\tilde{s}=s-i \ep$ and
\bea\label{twopointstress}
&&\la T_{\mu\nu}(u)T_{\alpha\beta}(v)\ra=\frac{d_T}{|u-v|^{2(d+1)}}S_{\mu\nu,\alpha\beta}(u-v),\nn\\
&&S_{\mu\nu,\alpha\beta}(u)=\frac{1}{2}\lb I_{\mu\alpha}(u)I_{\nu\beta}(u)+(\mu\leftrightarrow\nu)\rb-\frac{1}{d+1}\delta_{\mu\nu}\delta_{\alpha\beta}\nn\\
&&I_{\mu\nu}(u)=\delta_{\mu\nu}-\frac{2u_\mu u_\nu}{|u|^2}
\eea
Then,
\bea
&&\p_n A^{00}(n)\big|_{n\to 1}=\frac{d \mathcal{C}_d d_T}{d+1}\nn\\
&&\mathcal{C}_d=(-1)^d\int_{-\infty}^\infty \frac{ds}{(2\sinh(\tilde{s}/2))^{2(d+2)}}.
\eea
One can perform the integral explicitly
\bea
\mathcal{C}_d=\frac{2}{(d+2)}{}_2F_1\left[2(2+d),2+d,3+d,-1\right]=\frac{2}{(d+2)}\frac{\Gamma(d+3)^2}{\Gamma(5+2d)}\nn
\eea
Therefore,
\bea\label{CFTs2ndorder}
&&\delta^{(2)}S=-\frac{(d+1) \: \mathcal{C}_d d_T}{2d}\lb \frac{2l}{\lambda_T}\rb^{2(d+1)}\nn\\
&&=-\frac{2(d+1)^2\:\Gamma(d+3)\Gamma(d) d_T}{2\Gamma(5+2d)}\lb \frac{2l}{\lambda_T}\rb^{2(d+1)}\nn
\eea
Note that here $d_T= \la T_{00}T_{00}\ra (d+1)/d$, and in $d=1$ we have $d_{T}=c/(2\pi^2)$ therefore
\bea
&&\delta^{(2)}S=-\frac{4c}{15\pi^2}\lb\frac{l}{\lambda_T}\rb^{4}
\eea
which is the same as the result we found in two dimensions.

In $d>2$ we do not know the entanglement entropy in the reduced state of the Gibbs ensemble, $\rho^T_l$, however, if the theory is holographic we can compare the result with the prediction of the Ryu-Takayanagi formula.
Next, we show that the above result can be reproduced using a gravitational calculation in a black hole background. 
The calculation of the entanglement entropy of excited states in the small size limit and matching it with the black hole holographic entropy have appeared earlier in a very nice paper \cite{Sarosi:2016atx}\footnote{We thank G. Sarosi and T. Ugajin for bringing their paper to our attention.}. Instead of the Renyi entropy calculation presented here the authors use a  replica trick that directly computes the relative entropy \cite{Lashkari:2015dia}.

\subsection{Holographic theories}
Consider the thermal state of a holographic CFT in flat space dual to the planar black hole
\bea
ds^2=\frac{L^2}{z^2}\lb -f(z)dt^2+d\vec{x}_d^2+\frac{dz^2}{f(z)}\rb, \qquad f(z)=1-\frac{z^{d+1}}{z_h^{d+1}}.
\eea
Here, $z_h$ is related to the thermal wavelength $z_h=\frac{(d+1)\beta}{4\pi}$. The entanglement entropy of the reduced state on a ball of radius $l$ is the area of an extremal surface in the bulk anchoring on the boundary of the subsystem:
\bea
S(\rho_T,l)=\frac{L^dS_{d-1}}{4G}\int_0^l dr \frac{r^{d-1}}{z^d}\sqrt{1+\frac{(\p_r z)^2}{f(z)}}
\eea
 It is convenient to switch to the Fefferman-Graham coordinates to compute the entanglement entropy perturbatively in $l/\beta\sim l/z_h$:
\bea
&&ds^2=\frac{L^2}{z^2}(dz^2+g_{\mu\nu}(z,x^\mu)dx^\mu dx^\nu),\nn\\
&&g_{\mu\nu}(z,x^\mu)=\eta_{\mu\nu}+a z^{d+1} T_{\mu\nu}+ a^2 z^{2(d+1)}(n_1 T_{\mu\alpha}T^\alpha_{\:\nu}+n_2 \eta_{\mu\nu}T_{\alpha\beta}T^{\alpha\beta})+\cdots
\eea
where $a=\frac{16\pi G}{(d+1) L^d}$, $n_1=1/2$ and $n_2=-\frac{1}{8d}$. 
The bulk Ricci tensor written in these coordinates with $\rho=z^2/L^2$ (dimensionless) is
\bea
&&R_{\rho\rho}=-\frac{d}{\rho^2}-\frac{1}{2}g^{\mu\nu}g''_{\mu\nu}+\frac{1}{4}(g^{\mu\mu})^2(g'_{\mu\mu})^2\nn\\
&&L^2 R_{\mu\mu}=-2\rho g''_{\mu\mu}+2\rho g^{\mu\mu}(g'_{\mu\mu})^2-\rho g'_{\mu\mu}g^{\nu\nu}g'_{\nu\nu}+(d-2)g'_{\mu\mu}+g_{\mu\mu}g^{\nu\nu}g'_{\nu\nu}-\frac{d}{\rho}g_{\mu\mu}.\nn
\eea

Perturbatively in $l$ we find that the vacuum subtracted entropy is \cite{Blanco:2013joa}\footnote{Note that there is a typo in equation (3.55) of that paper.}
\bea\label{secondentropy}
&&\delta^{(1)}S=\frac{2\pi \omega_{d-1}T_{00} l^{d+1}}{d(d+2)}=\frac{2\pi \omega_{d-1} d_T}{d(d+2)}\lb\frac{l}{\lambda_T}\rb^{d+1}\nn\\
&&\delta^{(2)}S=-\frac{\pi^{3/2}(d+1)\omega_{d-1}\Gamma(d)}{2^{d+2}(d+2)\Gamma(d+5/2)}\lb\frac{8\pi G}{L^d}\rb T_{00}^2 l^{2(d+1)}\nn\\
&&=-\frac{\pi^{3/2}(d+1)\omega_{d-1}\Gamma(d)}{2^{d+2}(d+2)\Gamma(d+5/2)}\lb\frac{8\pi G d_T}{L^d}\rb d_T\lb\frac{l}{\lambda_T}\rb^{2(d+1)}
\eea
where we have used $T_{00}=\frac{d_T}{\lambda_T^{d+1}}$ and $\omega_d=\frac{2\pi^{(d+1)/2}}{\Gamma((d+1)/2)}$. The first term is simply the first law of entanglement entropy. The quantity $\frac{L^d}{8\pi G}$ is related to the two-point function of stress tensor as:
\bea\label{dTholo}
d_T=\frac{d+2}{d}\frac{\Gamma(d+2)}{\pi^{(d+1)/2}\Gamma((d+1)/2)}\frac{L^d}{8\pi G}.
\eea
Plugging this back in (\ref{secondentropy}) gives
\bea
\delta^{(2)}S=-\frac{(d+1)^2\Gamma(d+3)\Gamma(d)d_T}{\Gamma(2d+5)}\lb\frac{2l}{\lambda_T}\rb^{2(d+1)}
\eea
This is exactly the answer we found in the field theory in (\ref{CFTs2ndorder}) for the entanglement entropy of the universal density matrix in arbitrary dimension $d$. 

If the local ETH hypothesis is correct in holographic CFTs, the reduced density matrix in any energy eigenstate is well approximated by the ETH density matrix (\ref{lowtempd}). According to holography, the gravity dual of a heavy energy eigenstate is a black hole of the same energy density. Therefore, if the local ETH holds the entanglement entropy of the ETH density matrix should match the entanglement entropy computed holographically in the dual black hole geometry. In this section, we checked that in the same temperature limit $l/\lambda_T\ll 1$, indeed, the local ETH hypothesis passes this consistency check.

\section{Local equilibrium} \label{sec:gen}

Up to this point we were only concerned with the eigenstate thermalization hypothesis. We showed that the reduced density matrix of small subsystems in energy eigenstates are universal. Energy eigenstates are highly fine tuned and that their time-evolution is given by just an overall phase. 
Intuitively, we expect the density matrix of small subsystems to be only a function of energy not only in translationally-invariant energy eigenstates but also in all states that have spatial and time dependence over scalecs much larger than the size of the subsystem. In this section, we establish that this is indeed the case by studying the reduced density matrices in two classes of time-dependent states: ``coherent" states, and arbitrary superpositions of $N\ll e^{S(E)/2}$ energy eigenstates.
\subsection{Time-dependent coherent states}

We define ``coherent states" $|\Phi(\vec{s})\ra$ via a Euclidean path-integral with a local operator inserted at $\vec{s}$ inside the unit ball in the radial quantization frame:
\bea
&&|\Phi(\vec{s})\ra=e^{s^\mu P_\mu}\Phi(0)e^{-s^\mu P_\mu}|\Omega\ra
\eea
We can use the rotational symmetry of the unit ball to bring the operator insertion to the point $(r=e^\tau,\theta_1=\alpha)$ and $\theta_i=0$ for all $i>1$. Coherent states include a superposition of many energy eigenstates, and hence evolve non-trivially in time. Mapping to the Rindler space the operators that create and annihilate the state go to, respectively, $Y^\mu_-$ and $Y^\mu_+$:
\bea
(Y_\pm^0,Y_\pm^1)=\lb\frac{-\sin\theta_0\sinh\tau_\pm}{\cos(\theta_0+\alpha)-\cosh\tau_\pm},\frac{\cos\alpha-\cos\theta_0\cosh\tau_\pm}{\cos(\theta_0+\alpha)-\cosh\tau_\pm} \rb,\qquad \:Y_\pm^{i>1}=0
\eea
where $\tau_\pm=\pm\tau_0- i t$ and we have analytically continued to the real time to keep track of the time evolution of the state. 
The analytic continuation in time is achieved by treating $\tau_\pm$ as a real parameter. 

The parameter $\tau_0$ controls the width and angular dependence of the energy profile around $\mathbb{S}^d$ at time $t=0$. To see that we compute the energy density in this spinless primary state:
\bea
\la\phi_{\alpha,\tau_0}(t)|T_{00}(\theta,0,\cdots)|\phi_{\alpha,\tau_0}(t)\ra_{Cyl}&=&\frac{h_a}{L^{d+1}\omega_d}\left[ \frac{\sinh^2\lb\frac{\tau_--\tau_+}{2} \rb}{(\cos(\alpha-\theta)-\cosh\tau_-)(\cos(\alpha-\theta)-\cosh\tau_+)}\right]^{\frac{d+1}{2}}\nn\\
&=&\frac{h_a}{L^{d+1}\omega_d}\frac{1}{(\lb\cos t\coth\tau_0-\cos(\alpha-\theta)\csch\tau_0)^2+\sin^2t\rb^{(d+1)/2}}\nn
\eea 
At $t=0$ the energy density around $\mathbb{S}^d$ has its peak value $\coth^2(\tau_0/2)$ at the point $(\alpha,0,\cdots)$. In the thermodynamic limit of small subsystem $l/L\ll 1$ the energy density is constant over the subsystem
\bea
&&\ep(t,\theta\in B)=\frac{h_a}{L^{d+1}\omega_d \xi^{d+1}(t)}(1+O(1/L))\nn\\
&&\xi^2(t)=(\cos t\coth\tau_0-\cos\alpha\csch\tau_0)^2+\sin^2t\nn\\
&&(Y_\pm^0,Y_\pm^1)=\lb\frac{l\sinh\tau_\pm}{L (\cosh\tau_\pm-\cos\alpha)},1-\frac{l\sin\alpha}{L(\cosh\tau_\pm-\cos\alpha)} \rb
\eea
The ``local" length scale associated to the energy density is
\bea
\lambda_T(\tau_0,\alpha,t)=\xi(t) L \lb \frac{\omega_d d_T}{h_a} \rb^{\frac{1}{(d+1)}}
\eea

Then, the distance between the operator insertions is
\bea
|Y_+-Y_-|^2=\frac{4l^2}{L^2\xi(t)}
\eea
and the density matrix becomes
\bea
tr(\tilde{\psi}\cdots)=\sum_{p\in\mathcal{A}_{therm}}C^{p,\hat{n}}_{\phi,\phi}\xi^{-h_p}|\hat{n}|^{h_p}\mO_p^{\hat{n}}(Y_-)=\sum_{p\in\mathcal{A}_{therm}}f^{\hat{n}}_p (l/\tilde{\lambda}_T(t))^{h_p}\mO^{\hat{n}}_p(Y_-).
\eea
which shows that the reduced density matrix is universal with $\lambda_T$ multiplied by $\xi(t)$. That is to say at any time $t$ the reduced density matrix is in equilibrium with a time-dependent thermal wavelength $\xi(t)$.

\subsection{Arbitrary initial states}

An arbitrary CFT state in the Schrodinger picture expanded in the energy eigen-basis is
\bea
|\chi(t)\ra=\sum_{a=1}^N e^{i h_a t/L}c_a|\psi_a\ra
\eea
The reduced density matrix on a ball-shaped region in this state is a partial trace over the complement region
\bea
\rho_{B_R}(t)=tr_{B_R^c}|\chi(t)\ra\la\chi(t)|=\sum_{a b}c_ac_b^*\:e^{i t(h_a-h_b)}tr_{B_R^c}|\psi_a\ra\la \psi_b|
\eea
Now, it is straightforward to see
\bea
&&\|\rho_{B_R}(t)-\sum_a |c_a|^2 \rho_{uni}(E=E_a)\|\leq  sup_{a\neq b}\|\sigma_{ab}\|\Big |\sum_{a \neq b} c_a c_b^*\Big ||\nn\\
&&\leq \eta e^{-S(E)/2}(\sum_{a=1}^N |c_a|)^2\leq \eta N e^{-S(E)/2}
\eea
for some $\eta=O(1)$. 
Therefore, as long as the number of superposed energy eigenstates $N$ does not scale with entropy the reduced density matrix is well-approximated with a classical mixture of universal density matrices:
%
\bea
\int dE \: p(E) \rho_{uni}(E)
\eea
which does not evolve in time. If the state has $\la \chi(t)|H|\chi(t)\ra =E_0$ and $\la \chi(t)|H^2-E_0^2|\chi(t)\ra =\Delta E_0$ then the density matrix is approximately
\bea
\rho_{uni}(E_0)+\frac{\Delta E_0}{2}\p_E^2\rho_{uni}(E)|_{E_0}+\cdots
\eea

Quenching an energy eigenstate with a local operator of energy order one is an example of a state that necessarily includes a large number of energy eigenstates.

%
%
%
%

\section{Conclusions}
%
%

In this work, we continue the study of the Eigenstate Thermalization Hypothesis (ETH) in the context of Conformal Field Theories initiated in \cite{Lashkari:2016vgj}. In that paper, we formulated the {\it subsystem ETH} in CFTs as a statement about the smooth dependence of the reduced density matrix of an energy eigenstate on energy. We proved that if  ETH is satisfied at the level of individual local operators ({\it local ETH}), the subsystem ETH follows. 

In \cite{Lashkari:2016vgj} it was shown that the ETH density matrix exhibits a great degree of universality provided that the subsystem in question is small compared to the total volume. When the subsystem is small in comparison to the inverse effective temperature, the ETH density matrix admits a perturbative expansion in terms of the light primary operators \eqref{uni2}. In 2d CFTs the statement of ETH implies that no operator outside of the Virasoro descendants of identity contributes to the OPE of any two heavy Virasoro primaries. As a result the ETH density matrix exhibits a greater degree of universality, depending only on the effective temperature and the central charge, but on other detail of the underlying theory \eqref{psiETH2d}. 

In section \ref{sec:nee} of the paper we provided an argument based on the equivalence of ensembles, modified for the case of CFTs, to argue that the ETH density matrix for a small subsystem is trace-distance close to other thermal ensembles, the reduced canonical and the microcanonical ones. This general argument is further supported by the calculation and comparison of the eigenstate entanglement entropy with the holographic one in section \ref{sec:hd}.  In case of two dimensions, because of the additional conservation laws, the canonical ensemble must be substituted by the grand canonical ensemble that includes an infinite number of conserved KdV charges -- the Generalized Gibbs Ensemble. A new representation of the ETH density matrix and its equivalence with the thermal one in the limit of infinite central charge is demonstrated in section \ref{sec:2d}. There we also calculate the von Neumann and the Renyi entropies for the eigenstate and discuss the finite $c$ case. 

Finally, in section \ref{sec:gen} we discuss the reduced density matrix of time-dependent coherent states and show that their reduced density matrix on a small subsystem is well-described by the universal ETH density matrix with time-dependent effective temperature.

\section*{Acknowledgements}
We would like to thank John Cardy, Liam Fitzpatrick, Thomas Hartman, Matthew Headrick, Tarun Grover, Mark Srednicki, Matthew Walters and Sasha Zhiboedov for valuable discussions. The research of NL is supported in part by funds provided by MIT-Skoltech Initiative.  AD is supported by NSF grant PHY-1720374. This work is supported by the Office of High Energy Physics of U.S. Department of Energy under grant Contract Number DE-SC0012567.

\appendix

\section{Rindler space: a convenient conformal frame}\label{AppA}

Consider a $(d+1)$-dimensional CFT in radial quantization with a ball-shaped subsystem of angular size $\theta_0$ on $\mathbb{S}^d$ at $r=1$. According to the operator/state correspondence the density matrix in the subsystem is given by a path-integral over the $(d+1)$-dimensional space with two operators inserted, $\Psi$ at $r=\ep$ and $\Psi^\dagger$ at $r=1/\ep$ with $\ep\to 0$, and a cut open at the location of the subsystem. The initial metric in the radial quantization is 
\bea
ds^2=dr^2+r^2d\Omega_d^2
\eea
with $(\theta_1, \cdots \theta_d)$ the coordinates on $S^d$. We perform the following conformal transformation
\bea
&&\frac{L(r^2-1)}{1+r^2+2r\cos\theta_1}=\frac{X^0}{1-2 X^1+X\cdot X},\ \ \nn
\frac{2L r \sin\theta_1\cos\theta_2}{1+r^2+2r\cos\theta_1}=\frac{(1-X\cdot X)/2}{1-2 X^1+X\cdot X}, \\
&&\frac{2L r \sin\theta_1\sin\theta_2\cdots \cos\theta_{i+1}}{1+r^2+2r\cos\theta_1}=\frac{X^i}{1-2 X^1+X\cdot X},\quad d>i>1\nn\\
&&\frac{2L r \sin\theta_1\sin\theta_2\cdots \sin\theta_d}{1+r^2+2r\cos\theta_1}=\frac{X^d}{1-2 X^1+X\cdot X},\nn \ \ 
L=\frac{1}{2}\cot(\theta_0/2).
\eea
that maps the subsystem at $r=0$ and $\theta_1\leq \theta_0$ to the negative half-space, i.e. $(0,X^1<0,0\cdots 0)$. Here $L$ is the radius of $\mathbb{S}^d$ in units where $R$ is set to one. The new metric in the $X$-coordinates that we call Rindler frame is given by
\bea
&&ds^2=\Lambda(X)^2 dX^i dX^i\nn\\
&&\Lambda(X)=\lb X^0-\frac{L V_-}{2}-\frac{V_+}{8L}\rb^{-1}\nn\\
&&V_\pm=(1\pm 2X^1+X\cdot X).
\eea

 In these coordinates the path-integral without operator insertions prepares the Rindler density matrix in vacuum. The operators $\Psi$ and $\Psi^\dagger$ are now inserted at $X_-$ and $X_+$ respectively.
 \bea
 &&X_\pm=(\pm \sin\theta_0,\cos\theta_0,0\cdots, 0),\nn\\
 &&\Lambda(X_-)=(2\sin\theta_0)^{-1},\nn\\
 &&\Lambda(X_+)=\ep^{-2}(2\sin\theta_0)^{-1}.
 \eea
 Under this map a conformal primary transforms according to
 \bea
\la  \Psi(r=0)\cdots \ra_{\Lambda(X)\delta_{ij}}=\Lambda(X(r=0))^{-h}\la \Psi(X(r=0)\cdots \ra_{\delta_{ij}}\nn
 \eea
Therefore,
\bea
\la\Psi(1/\ep)\Psi(\ep)\cdots \ra_{radial}=(2\ep\sin\theta_0)^{2h}\la \Psi(X_+)\Psi(X_-)\cdots \ra_{Rind}\nn
\eea
 In the thermodynamic limit $\theta_0\ll 1$ the distance between $\Psi$ and $\Psi^\dagger$ goes to zero: $|X_+-X_-|=2\sin\theta_0\ll 1$, and we use the OPE to obtain
 \bea
 &&\la\Psi(1/\ep)\Psi(\ep)\cdots \ra_{radial}=\ep^{2h}\sum_p C^p_{\psi\psi} (2\sin\theta_0)^{h_p}\la \mO_p(X_0)\cdots\ra\nn.
 \eea

\section{Global descendants in two dimensions}\label{AppB}
Consider the OPE of two quasi-primaries $\Psi$ in $CFT_2$
\bea
\frac{\Psi(z,\bar{z})\Psi(0,0)}{\la\Psi(z,\bar{z})\Psi(0,0)\ra}=\sum_p C^p_{\psi\psi}\sum_{j,\bar{j}}\frac{a_{\psi\psi p}^j\bar{a}_{\psi\psi p}^{\bar{j}}}{j! \bar{j}!} z^{h_p+j}\bar{z}^{\bar{h}_p+\bar{j}}\p^j\bar{\p}^{\bar{j}}\Phi_p
\eea
where $\Phi_p$ are quasi-primaries and
\bea
&&a^j_{\psi\psi p}=\frac{C(j,h_p+j-1)}{C(j,2h_p+j-1)},\qquad \bar{a}^j_{\psi\psi p}=\frac{C(\bar{j},\bar{h}_p+\bar{j}-1)}{C(\bar{j},2h_p+\bar{j}-1)}\nn\\
&&C^p_{\psi\psi}=\frac{1}{\la \Phi_p\Phi_p\ra } \la\psi|\Phi_p|\psi\ra, \qquad C(j,h)=\frac{\Gamma(h+1)}{\Gamma(j+1)\Gamma(h-j+1)}
\eea
In the thermodynamic limit $z=l/L$, $h$ and $L$ go to infinity with $\lambda_T\sim L/\sqrt{h}$ kept fixed we have
\bea
a^j_{\psi\psi p} z^j\to 0\qquad \forall j>0.
\eea
Therefore, all the derivative terms are subleading, and we have
\bea
\frac{\Psi(z,\bar{z})\Psi(0,0)}{\la\Psi(z,\bar{z})\Psi(0,0)\ra}=\sum_p C^p_{\psi\psi}z^{h_p}\bar{z}^{\bar{h}_p}\Phi_p+O(1/L).
\eea

This argument generalizes to higher dimensions. Consider a primary $\mO_p$ and its first descendant. Then, the OPE coefficients are the same order 
\bea
\frac{C^{\p\mO_p}_{\psi\psi}}{C^{\mO_p}_{\psi\psi}}=\frac{d_{\mO_p}}{d_{\p\mO_p}}\frac{\la \Psi(\infty) \p\mO_{p}(1)\Psi(0)\ra}{\la \Psi(\infty) \mO_{p}(1)\Psi(0)\ra}=2h_p(2h_p-1)h_p=O(h_\psi^0)
\eea
however, by in the OPE of $\Psi$s, the derivative term has an extra power of $l/L$ and is hence more suppressed.

\section{Thermodynamically relevant quasi-primaries}\label{AppC}
In this appendix, we expand the reduced state on an interval of length $2k$ in a highly excited primary energy eigenstate, and find the quasi-primaries that contribute to the universal density matrix, that are $\mT_{2k}$ in (\ref{psiuniv2}).
Consider a primary energy eigenstate $|\psi_a\ra$ and its correponding operator $\Psi_a$. In Rindler coordinates, the density matrix is created by the insertion of operator 
\bea\label{LkOPE}
\frac{\Psi_a(z,\bar{z})\Psi_a(0)}{\la\Psi_a(z,\bar{z})\Psi_a(0)\ra}=\sum_p \sum_{\{k,\bar{k}\}}C^{p\{k,\bar{k}\}}_{aa}z^{h_p+K}\bar{z}^{\bar{h}_p+\bar{K}}L_{-\{k\}}\bar{L}_{-\{\bar{k}\}}\mO_p
\eea
in the Euclidean path-integral. Here, $\{k\}=\{k_1\cdots k_l\}$, $K=k_1+\cdots +k_l$, and $z=x/L$ with $L$ going to infinity in the thermodynamic limit. The OPE coefficient $C^{p,\{k,\bar{k}\}}_{aa}$ (growing with $h_a$) competes with the vanishing coefficient $(x/L)^{h_p+K}$.

To determine what operator survive the thermodynamic limit in (\ref{LkOPE}) we need to investigate the growth of this OPE coefficient with $h_a$. It is convenient to define the OPE coefficient with lowered indices \cite{polchinski1998string}
\bea\label{OPEcoeffformula}
&&C^{p,\{k,\bar{k}\}}_{aa}=\sum_{\{k',\bar{k'}\}}\left[\mathcal{M}^{-1}\right]^{p\{k\}\{k'\}}\left[\mathcal{M}^{-1}\right]^{p,\{\bar{k}\}\{\bar{k'}\}}C_{aa,p\{k'\}\{\bar{k'}\}}\nn\\
&&C_{aa,p\{k'\}\{\bar{k'}\}}=\mathcal{L}_{-\{k'\}}\bar{\mathcal{L}}_{-\{\bar{k'}\}}\la\Psi_a(\infty)\Psi_a(1)\mO_p(y)\ra\Big|_{y=0}.
\eea
The matrix $\mathcal{M}$ is the Kac matrix defined by $\mathcal{M}_{\{k\},\{k'\}}(h_p,c)=\la h_p|L_{\{k\}}L_{-\{k'\}}|h_p\ra$, and is independent of $h_a$. We only need to consider $C_{aa,p\{k\}\{\bar{k}\}}$. The differential operator $\mathcal{L}_{-\{k\}}\equiv\mathcal{L}_{-k_1}\cdots \mathcal{L}_{-k_l}$ with each $\mathcal{L}_{-k}$ acting as
\bea\label{OPEquasi1}
&&\mathcal{L}_{-k}\la \Psi_a(\infty)\Psi_a(1)\mO_p(y)\ra=\nn\\
&&C^p_{aa}\lim_{(z,\omega)\to(\infty,1)}z^{2h_a}\lb h_a(k-1)(z^{-k}+\omega^{-k})-(z^{1-k}\p_z+\omega^{1-k}\p_\omega)\rb(z-\omega)^{h_p-2h_a}(z\omega)^{-h_p}\nn\\
&&=C^p_{aa}(h_a(k-1)+h_p)\simeq C^p_{aa}h_a(k-1)\ .
\eea
At order $K$ we are comparing OPE coefficients of operators of the form $L_{k_1}L_{k_2}\cdots L_{k_l}\mO_p$ with $k_1+\cdots+k_l=K$. From (\ref{OPEquasi1}) it is clear that the OPE coefficient of operators with $L_{-1}$ does not grow fast enough with $h_a$ and they drop out of the thermodynamic limit, which is consistent with the result in appendix \ref{AppB}. We only need to consider the case with $k_i>1$. Then,
\bea
C_{aa,p\{k_1,\cdots,k_l\}\{\bar{k}_1,\cdots \bar{k}_m\}}\sim h_a^{l+m}.
\eea
For even $K$ the OPE coefficient of the quasi-primary that includes $L_{-2}^{K/2}$ wins over other terms. When $K$ is odd none of the OPE coefficients are large enough to compete with $(x/L)^{K+\bar{K}}$. 
Therefore, the sum over $\{k',\bar{k}'\}$ in (\ref{OPEcoeffformula}) only has one term, and
\bea
&&C^{p\{k,\bar{k}\}}_{aa}=C^p_{aa}b^{p,\{k,\bar{k}\}}h_a^{K/2+\bar{K}/2}\nn\\
&&b^{p,\{k,\bar{k}\}}=\left[\mathcal{M}^{-1}\right]^{\{2,\cdots,2\}\{k\}}\left[\mathcal{M}^{-1}\right]^{\{\bar{2},\cdots,\bar{2}\}\{\bar{k}\}}
\eea
where $K$ and $\bar{K}$ are both even. Note that in two dimensions $C^p_{aa}=0$ for all non-identity Virasoro primaries $p$. Therefore,
\bea\label{OPE2dlargeh}
&&\frac{\Psi_a(e^{i\theta_0},e^{-i\theta_0})\Psi_a(e^{-i\theta_0},e^{i\theta_0})}{\la\Psi_a(e^{-i\theta_0},e^{i\theta})\Psi_a(e^{-i\theta_0},e^{i\theta_0})\ra}= \lb\sum_{\{k\}}b^{\{k\}}(2\sqrt{h}_a \sin\theta_0)^K L_{-\{k\}}\rb\times h.c.\nn\\
&&=\lb\sum_{m\in\mathbb{N}}(2\sqrt{h}_a \sin\theta_0)^{2m}\sum_{k_1+\cdots k_l=2m}[\mathcal{M}^{-1}]^{\{2\cdots 2\}\{k_1\cdots k_l\}} L_{-k_1}\cdots L_{-k_l}\rb\times h.c.\nn\\
&&=\lb \sum_{m\in\mathbb{N}}\lb\frac{2l}{\sqrt{2\pi}\lambda_T}\rb^{2m} \frac{c^m}{d_{2m}}\mT^{(0)}_{2m}\rb\times h.c.\nn\\
&&\frac{1}{d_{2m}}\mT^{(0)}_{2m}\equiv\sum_{k_1+\cdots k_l=2m}[\mathcal{M}^{-1}]^{\{2\cdots 2\}\{k_1\cdots k_l\}} L_{-k_1}\cdots L_{-k_l}
\eea
where in the last two lines we have defined an operator $\mT^{(0)}_{2m}$ with the norm $d_{2m}=\la \mT^{(0)}_{2m}(1)\mT^{(0)}_{2m}(0)\ra$.
The first few $\mT^{(0)}_{2k}$ are 
\bea\label{tKdirect}
&&\mT^{(0)}_{2}=L_{-2},\qquad \mT^{(0)}_{4}= L_{-2}^2-\frac{3}{5}L_{-4}\nn\\
&&\mT^{(0)}_{6}=L_{-2}^3+\frac{93}{70c+29}L_{-3}^2-\frac{3(42c+67)}{70c+29}L_{-4}L_{-2}-\frac{6(10c+13)}{70c+29}L_{-6}\nn\\
&&\mT_8^{(0)}=L_{-2}^4+
-\frac{6 \left(630 c^2+3471 c-557\right) L_{-4}L_{-2}^2}{5 c (210 c+661)-251}+\frac{(5844-1512 c) L_{-5}L_{-3}}{5 c (210 c+661)-251}+\nn\\
&&\frac{27 (c (42 c+265)-167) L_{-4}^2}{5 c (210 c+661)-251}-\frac{24 (c (150 c+569)+67) L_{-6}L_{-2}}{5 c (210 c+661)-251}-\frac{6 (5 c (126 c+463)-543) L_{-8}}{5 c (210 c+661)-251}\nn\\
&&\mT^{(0)}_{10}=L_{-2}^5-\frac{12 \left(8250 c^2+58115 c-7161\right) L_{-6}L_{-2}^2}{25 c (462 c+3067)+3767}+\left(-\frac{12 (11650 c+15341)}{25 c (462 c+3067)+3767}-18\right) L_{-8}L_{-2}\nn\\
&&+\frac{36 (4358-3225 c) L_{-7}L_{-3}}{25 c (462 c+3067)+3767}+\frac{36 (c (1650 c+16783)-8405) L_{-6}L_{-4}}{25 c (462 c+3067)+3767}+\frac{(31032 c+220236) L_{-5}^2}{25 c (462 c+3067)+3767}\nn\\
&&+\frac{9 (45 c (154 c+1873)+25133) L_{-4}^2L_{-2}}{25 c (462 c+3067)+3767}+\left(-\frac{48 (5115 c+1081)}{25 c (462 c+3067)+3767}-6\right) L_{-4}L_{-2}^3\nn\\
&&+\frac{30 (5115 c+1081) L_{-3}^2L_{-2}^2}{25 c (462 c+3067)+3767}-\frac{924 (90 c+259) L_{-5}L_{-3}L_{-2}}{25 c (462 c+3067)+3767}-\frac{18 (5115 c+1081) L_{-4}L_{-3}^2}{25 c (462 c+3067)+3767}\nn\\
&&-\frac{504 (c (300 c+1693)+266) L_{-10}}{25 c (462 c+3067)+3767}\nn\\
&&d_2=\frac{c}{2},\qquad d_4=\frac{c(5c+22)}{10}, \qquad d_6=\frac{3c(2c-1)(5c+22)(7c+68)}{4(70c+29)}\\
&&d_8=\frac{3 c (2 c-1) (3 c+46) (5 c+3) (5 c+22) (7 c+68)}{10 c (210 c+661)-502}\nn\\
&&d_{10}=\frac{15 c (2 c-1) (3 c+46) (5 c+3) (5 c+22) (7 c+68) (11 c+232)}{4 (25 c (462 c+3067)+3767)}
\eea
Note that 
\bea
&&(L_{-n-2}\Phi)(\omega)=\frac{1}{n!}\p^nT(\omega),\nn\\
&&L_{-2}^3(\omega)=(T(T T))(\omega),\qquad L_{-3}^2(\omega)=(\p T\p T)(\omega),\nn\\
&&\lb L_{-3}^2+L_{-4}L_{-2}+L_{-2}L_{-4}\rb(\omega)=\frac{1}{2}\p^2(T T)(\omega).
\eea
Then, we find
\bea
&&\mT^{(0)}_{2}=T,\qquad \mT^{(0)}_{4}=(T T)-\frac{3}{10}\p^2T\nn\\
&&\mT^{(0)}_{6}=(T(T T))+\frac{9(14c+43)}{2(70c+29)}(\p T\p T)-\frac{3(42c+67)}{4(70c+29)}\p^2(T T)-\frac{(22c+41)}{8(70c+29)}\p^4T\nn.
\eea

An alternative way to construct the quasi-primary operators $\mT_{2k}$ is by choosing the basis where the Kac matrix is diagonal.
 In this basis, it is evident that the only quasi-primaries that include the term $L_{-2}^{m}=T^{m}(0)$ propagate. Here, $T^m=(T(T(T\cdots T)))$. We can choose our operator basis such that at even order $K$ only one quasi-primary includes $L_{-2}^{K/2}$ which becomes our operator of interest $\mT^{(0)}_{2k}$. Below, we describe how to construct it at any even order $K$.


\begin{enumerate}
\item Consider an arbitrary superposition of $L_{-\{k\}}$ with no $L_{-2,\cdots -2}$ : $\sum_{\{k\}\neq (2,\cdots,2)}a_k L_{-\{k\}}(0)$.
\item Choose $a_{\{k\}}$ such that this state is annihilated by $L_1$. The result is the most generic quasi-primary with no $L_{-2,\cdots -2}$.
\item Find an arbitrary superposition state with $L_{-2,\cdots -2}$ that is perpendicular to the state above, and demand that it is killed by $L_1$. The resulting state is $\mT^{(0)}_{K}$.
\end{enumerate}

We end this appendix by consider the quasi-primaries $\mT_{2k}$ in the limit $h\gg c\gg 1$. In this limit, the expressions for the first $\mT_{2k}$ simplify to 
\bea
&&\frac{1}{d_{2m}}=\frac{1}{m!}\lb\frac{2}{c}\rb^m \nn\\
&&\mT_2^{(0)}=L_{-2},\qquad \mT_4^{(0)}=L_{-2}^2-\frac{3}{5}L_{-4},\nn\\
&&\mT_6^{(0)}=L_{-2}^3- \frac{9}{5} L_{-4}L_{-2}-\frac{6}{7} L_{-6}\nn\\
&&\mT_8^{(0)}=L_{-2}^4-\frac{24}{7} L_{-6}L_{-2}+\frac{27}{25} L_{-4}^2-\frac{18}{5} L_{-4}L_{-2}^2-\frac{18}{5} L_{-8}\nn\\
&&\mT_{10}^{(0)}=L_{-2}^5-18 L_{-8}L_{-2}+\frac{36}{7} L_{-6}L_{-4}-\frac{60}{7} L_{-6}L_{-2}^2+\frac{27}{5} L_{-4}^2L_{-2}-6 L_{-4}L_{-2}^3-\frac{144}{11} L_{-10}\ .
\eea
Therefore, the holomorphic part of the density matrix operator becomes
\bea\label{holouniv}
\sum_{m\in\mathbb{N}}\lb\frac{4l^2}{\pi\lambda_T^2}\rb^m\frac{1}{m!}\mT^{(0)}_{2m}=\sum_{m\in\mathbb{N}} \lb \frac{4\pi^2 l^2}{3\beta^2}\rb^m\frac{1}{m!} \mT^{(0)}_{2m}
\eea
It is convenient to write the universal density matrix in an exponentiated form in this limit:
\bea
&&\exp\lb\sum_{0<m\in\mathbb{N}} a_{2m}\lb\frac{2l}{\sqrt{\pi}\lambda_T} \rb^{2m} L_{-2m}\rb\nn\\
&&a_2=1,\qquad a_4=-\frac{3}{10},\qquad a_6=\frac{11}{70},\qquad a_8=-\frac{9}{140},\qquad a_{10}=-\frac{34}{1925},\qquad\cdots \ .\nn
\eea

\section{One-point functions on a torus}\label{AppL}

The finite temperature expectation value of a primary operator at finite volume in two-dimensions is a one-point function on a torus with modular parameter $\tau=\frac{i\beta}{L}$ where $L$ and $\beta$ are the periodicities of the spatial and time circles, respectively.  
Modular invariance related the one-point function at high temperatures to low temperatures
\bea
\la\mO_{p}\ra_{-1/\tau}=(-1)^{h_p-\bar{h}_p}\tau^{h_p}\bar{\tau}^{\bar{h}_p}\la \mO_p\ra_\tau.
\eea
Therefore, for $\tilde{\beta}=\frac{L}{\beta}$ we have
\bea
tr(\rho_\beta \mO_p)=\lb L T\rb^{2h_p}tr(\rho_{\tilde{\beta}}\mO)
\eea

%
The parameter $q=e^{2\pi i \tau}$ at $\tau=i L T$ becomes $q=e^{-2\pi L T}$ and small at  large $L T$. Therefore, we can expand the one-point function perturbatively in small $q$:
\bea
\la\mO_p\ra_q=\sum_{h,\bar{h}}C^p_{(h,\bar{h})(h,\bar{h})}\:q^{h-\frac{c-1}{24}}\bar{q}^{\bar{h}-\frac{c-1}{24}}\frac{1}{\eta(q)\eta(\bar{q})}\sum_{N=0}^\infty q^N H_{N,h,p}.
\eea
The coefficients $H_N$ are found using a recursive relation with the first term $H_{0,h,p}=1$ \cite{Hadasz:2009db}. At large $LT$ only the lowest dimension primary of dimension $(\Delta,\bar{\Delta})$ contributes
\bea
tr(\rho_{\tilde{\beta}} \mO_p)\simeq\frac{\sum_{h,\bar{h}}C^p_{(h,\bar{h})(h,\bar{h})}\:e^{-2\pi L T(h+\bar{h}-c/12)}}{\sum_{h,\bar{h}}\:e^{-2\pi L T(h+\bar{h}-c/12)}}\simeq C_{\Delta,\bar{\Delta}}^p\:e^{-2\pi L T(\Delta+\bar{\Delta})},
\eea
This conclude our estimate of the size of one-point function probes in the thermodynamic limit
\bea
tr(\rho_\beta \mO_p)=(TL)^{2h_p}e^{-2\pi L  T(\Delta+\bar{\Delta})}C^p_{\Delta,\bar{\Delta}}.
\eea
As expected in the limit $L T\to \infty$ the thermal one-point functions are exponentially suppressed. 

\section{Perturbative Renyi entropies}\label{AppD}
In this appendix, we compute the Renyi entropies of the universal density matrix $\psi$ via a direct calculation of $tr(\psi^n)$. 
We take the subsystem to have size $2x$, and the length scale associated with the energy density in $\psi$ to be $\lambda_T$. The trace of $\psi^n$ is computed by sewing $n$ copies of the path-integrals that prepares $\psi$ (the path-integral in Rindler space with the operator (\ref{psiETH2d}) on each copy). Therefore, the vacuum subtracted Renyi entropy of $\psi$ is
\bea\label{deltaSn}
&&\Delta S_n(\psi,x)=\frac{(n+1)c}{12n\pi }(2x/\lambda_T)^2+\nn\\
&&\frac{1}{1-n}\log\Big\la\prod_{j=1}^n\sum_{K_j\bar{K}_j}\lb\frac{2x\sqrt{c}}{\sqrt{2\pi}n\lambda_T}\rb^{K_j+\bar{K}_j} e^{2\pi i j (K_j-\bar{K}_j)/n}\frac{\mT_{K_j}(e^{2\pi i j/n})\mT_{\bar{K}_j}(e^{-2\pi i j/n})}{d_{K_j}d_{\bar{K}_j}}\Big\ra\nn \ .
\eea
We expand the above expression in powers of $2x/\lambda_T$ and consider the first few terms. The first term corresponds to $(K_j,\bar{K}_j)=(0,0)$ for all $j$ except for $K_0$ and $\bar{K}_0$. This term is equal to one by the normalization of two point functions.
The first non-trivial term appears at $j=2$ and $(2x/\lambda_T)^4$:
\bea
&&\frac{n}{2}\sum_{l=1}^{n-1}\sum_{\substack{
  K_1K_2=0,2,4\\\bar{K}_1\bar{K}_2=\bar{0},\bar{2},\bar{4}}}  \lb\frac{2x\sqrt{c}}{\sqrt{2\pi}n\lambda_T}\rb^{K_1+\bar{K}_1+K_2+\bar{K}_2} \frac{e^{2\pi i l/n(K_2-\bar{K}_2)}}{d^2_{K_1}d^2_{K_2}}\la (\mT_{K_1}\mT_{\bar{K}_1})(1)(\mT_{K_2}\mT_{\bar{K}_2})(e^{2\pi i l/n})\ra\nn\\
&&=\frac{n}{2}\sum_{l=1}^{n-1}\sum_{K,\bar{K}} \lb\frac{2x\sqrt{c}}{2\sqrt{2\pi}n\lambda_T}\rb^{2K+2\bar{K}}\frac{\sin(\pi l/n)^{-2(K+\bar{K})}}{d_Kd_{\bar{K}}}=\frac{c(x/\lambda_T)^4}{16\pi^2}\frac{(n^2-1)(n^2+11)}{90n^3}\nn.
\eea
At $j=3$ we have 6-point functions of $\Psi_a$ (3-point functions of $\mT_K$) 
\bea
&&\sum_{1\leq l<m<q\leq n-1}\sum_{K_1K_2K_3=2} \lb\frac{2x\sqrt{c}}{\sqrt{2\pi}n\lambda_T}\rb^{\sum_{i=1}^3(K_i+\bar{K}_i)} \frac{e^{2\pi i(l \delta K_1+m \delta K_2+q \delta K_3)/n}}{d_{K_1}^2d_{K_2}^2d_{K_3}^2}\times\nn\\
&&\Big\la \mT_{K_1}(e^{2\pi i l/n})\mT_{K_2}(e^{2\pi i m/n})\mT_{K_3}(e^{2\pi i q/n}) \mT_{\bar{K}_1}(e^{-2\pi i l/n})\mT_{\bar{K}_2}(e^{-2\pi i m/n})\mT_{\bar{K}_3}(e^{-2\pi i q/n})\Big\ra\nn\\
&&=\sum_{1\leq l<m<q\leq n-1} \frac{1}{(8d_2)^3}\lb\frac{2x\sqrt{c}}{n\sqrt{\pi}\lambda_T}\rb^6\frac{2C_{TTT}}{s_{lm}^2s^2_{mq}s_{ql}^2}
=(2x/\lambda_T)^6\frac{c}{32\pi^3}\frac{(n^2-1)(n^4-4)(n^2+47)}{2835n^5}\nn
\eea
where $\delta K_i=K_i-\bar{K}_i$ and $s_{lm}=\sin(\pi(l-m)/n)$. We have used the summation identities in \cite{Chen:2013kpa}. It is important to note that up to the order $(l/\lambda_T)^6$ the density matrix depends only on the energy density of the pure state.

Therefore, to the sixth order we find 
\bea
&&\Delta S_n(\psi,x)=\frac{(1+n)c}{12n\pi}(2x/\lambda_T)^2-\frac{(1+n)c}{120 n\pi^2}\frac{(n^2+11)}{12n^2}(2x/\lambda_T)^4\nn\\
&&+\frac{(1+n)c}{630n\pi^3}\frac{(4-n^2)(n^2+47)}{144n^4}(2x/\lambda_T)^6
\eea
The next non-trivial one-point function $\la\psi|\mT_4|\psi\ra$ contributes to the entanglement entropy at order $(l/\lambda_T)^8$. In the next appendix, we result above to the sixth order and compute the eighth-order term using the twist operator method.

\section{Twist operators}\label{AppE}
The correlation function (\ref{deltaSn}) that appears in the calculation of the Renyi entropy of the universal density matrix is $\mathbb{Z}_n$ symmetric. That is to say that it is invariant under $z\to e^{2\pi i/n}z$. 
An alternative way to compute this correlator is by employing twist operators in a $Z_n$-orbifold theory. Here, we use the orbifold theory to reproduce the result of the last subsection and extend it to the eighth order in subsystem size. In the orbifold theory, the vaccum-subtracted Renyi entropy in terms of the four-point function below
\bea\label{sigmasigma}
&&\Delta S_n(\psi,x)=\frac{1}{1-n}\log G_4(z,\bar{z}),\nn\\
&& G_4(z,\bar{z})=\frac{\la\Psi^n(\infty)\sigma_n(z)\sigma_n(1)\Psi^n(0)\ra}{\la \Psi(\infty)\Psi(0)\ra^n\la \sigma_n(z)\sigma_n(0)\ra}
\eea
where $z=e^{i x/L}$. The quasi-primaries of the orbifold theory take the form $\prod_{i=1}^n\mO^{(i)}$, where $\mO_i$ is the primary on the $i^{th}$ copy. Local ETH implies that this correlator is dominated by the Virasoro identity block. Below we use perturbation theory to compute Renyi entropies order by order in $2x/\lambda_T$. 

The quasi-primaries that contribute to the Virasoro identity block at even orders up to $z^6$ are
\bea
\text{order}\: z^2\qquad &&T^{(j)} \nn\\
\text{order}\: z^4\qquad &&T^{(i)}T^{(j)} (i\neq j),\quad \mT_4^{(j)}\nn\\ 
\text{order}\: z^6\qquad &&T^{(i)}T^{(j)}T^{(l)} (i\neq j\neq l\neq i),\quad \mT_4^{(j)}T_2^{(l)} (j\neq l),\quad \mT_6^{(j)}\nn\\
\text{order}\: z^8\qquad &&T^{(i)}T^{(j)}T^{(l)} T^{(m)}(\neq),\quad T^{(i)}T^{(j)}\mT^{(l)}_4 (\neq)\nn\\
&& \mT_4^{(j)}\mT_4^{(l)} (j\neq l),\quad \mT_6^{(j)}T^{(l)} (j\neq l),\quad \mT_8{(j)}\nn
\eea
where the symbol $\neq$ means that all pairs of indices are unequal.
These operators are listed in \cite{Chen:2013kpa}. The correlator factorizes into the holomorphic and anti-holomorphic parts
\bea
G_4(z,\bar{z})=|F(z,n,c)|^2
\eea
where the vacuum conformal block $F$ is only a function of cross ratio $z$, Renyi index $n$ and central charge $c$. 
\bea
&&F(z)=1+\sum_{ordered} \lb C^{T^{(j)}}_{\sigma_n\sigma_n}C^{T^{(j)}}_{\psi^n\psi^n}(1-z)^2+\lb C^{T^{(j)}T^{(l)}}_{\psi^n\psi^n}C^{T^{(j)}T^{(l)}}_{\sigma_n\sigma_n}+C^{\mT_4^{(j)}}_{\sigma_n\sigma_n}C^{\mT_4^{(j)}}_{\psi^n\psi^n}\rb (1-z)^4 \right.\nn\\
&&\left.+ \lb C^{T^{(j)}T^{(l)}T^{(q)}}_{\psi^n\psi^n}C^{T^{(j)}T^{(l)}T^{(q)}}_{\sigma_n\sigma_n}+2C^{\mT_4^{(j)}T^{(l)}}_{\psi^n\psi^n }C^{\mT_4^{(j)}T^{(l)}}_{\sigma_n\sigma_n}+C^{\mT_6^{(j)}}_{\sigma_n\sigma_n}C^{\mT_6^{(j)}}_{\psi^n\psi^n}\rb(1-z)^6\right.\nn\\
&&\left.+\lb C^{TTTT}_{\psi^n\psi^n}C^{TTTT}_{\sigma_n\sigma_n}+3C^{TT\mT_4}_{\psi^n\psi^n}C^{TT\mT_4}_{\sigma_n\sigma_n}+C^{\mT_4\mT_4}_{\psi^n\psi^n}C^{\mT_4\mT_4}_{\sigma_n\sigma_n}+2C^{\mT_6T}_{\psi^n\psi^n}C^{\mT_6T}_{\sigma_n\sigma_n}+C^{\mT_8}_{\psi^n\psi^n}C^{\mT_8}_{\sigma_n\sigma_n}\rb (1-z)^8\right.\nn\\
&&\cdots
\eea
where $\sum_{ordered}$ runs over all indices of the operator as $1\leq j_1<j_2<\cdots <j_k\leq n$.
At large $h$ we have $C_{\psi^n\psi^n}^{\mT_{k_1}\cdots \mT_{k_m}}=h^{k_1+\cdots k_m}$, and define $b_{\mT_{k_1}\cdots \mT_{k_m}}=\sum_{ordered} C^{\mT_{k_1}\cdots \mT_{k_m}}_{\sigma_n\sigma_n}$. These sums are computed in \cite{He:2017vyf}:
\bea
&&b_T=\frac{n^2-1}{12 n},\quad b_{\mT_4}=\frac{\left(n^2-1\right)^2}{288 n^3},\quad b_{\mT_6}=\frac{\left(n^2-1\right)^3}{10368 n^5},\quad b_{\mT_8}=\frac{\left(n^2-1\right)^4}{497664 n^7}\nn\\
&& b_{TT}=\frac{\left(n^2-1\right) \left(5 c (n+1) (n-1)^2+2 n^2+22\right)}{1440 c n^3},\nn\\
&& b_{T\mT_4}=\frac{\left(n^2-1\right)^2 \left(5 c (n+1) (n-1)^2+4 n^2+44\right)}{17280 c n^5}\nn\\
&&b_{T\mT_6}=\frac{\left(n^2-1\right)^3 \left(5 c (n+1) (n-1)^2+6 n^2+66\right)}{622080 c n^7}\nn\\
&&b_{\mT_4\mT_4}=\frac{1}{5806080 c (5 c+22) n^7}\lb 175 c^2 (n+1)^4 (n-1)^5\right.\nn\\
&&\left.+70 c \left(n^2-1\right)^3 \left(11 n^3-7 n^2-11 n+55\right)+8 \left(n^2-1\right) \left(n^2+11\right) \left(157 n^4-298 n^2+381\right)\rb\nn\\
&&b_{TTT}=\frac{(n-2) \left(n^2-1\right) \left(35 c^2 (n+1)^2 (n-1)^3+42 c \left(n^4+10 n^2-11\right)-16 (n+2) \left(n^2+47\right)\right)}{362880 c^2 n^5}\nn\\
&&b_{TT\mT_4}=\frac{(n-2) \left(n^2-1\right)}{14515200 c^2 n^7} \left(175 c^2 (n+1)^3 (n-1)^4+350 c \left(n^2-1\right)^2 \left(n^2+11\right)\right.\nn\\
&&\left.-128 (n+2) \left(n^4+50 n^2-111\right)\right)\nn\\
&&b_{TTTT}=\frac{(n-3) (n-2) \left(n^2-1\right)}{87091200 c^3 n^7} \left(175 c^3 (n+1)^3 (n-1)^4+420 c^2 \left(n^2-1\right)^2 \left(n^2+11\right)\right.\nn\\
&&\left.-4 c \left(59 n^5+121 n^4+3170 n^3+6550 n^2-6829 n-11711\right)+192 (n+2) (n+3) \left(n^2+119\right)\right)\nn
   \eea
Performing the $Z_n$ sums over trigonometric functions we find
\bea
&&F(z)=1+a_2 h(1-z)^2+a_4 h^2(1-z)^4+a_6 h^3(1-z)^6+\cdots\nn\\
&&a_2=\frac{(n^2-1)}{12n},\qquad a_4=\frac{(n^2-1)^2}{288n^2}+\frac{(n^2-1)(n^2+11)}{720n^3c}\nn\\
&&a_6=\frac{(n^2-1)^3}{10368n^3}+\frac{(n^2-1)^2(n^2+11)}{8640n^4c}+\frac{(n^2-1)(4-n^2)(n^2+47)}{22680n^5c^2}\nn\\
&&a_8=\frac{\left(n^3-3 n+3\right) \left(n^2-1\right)^4}{497664 n^7}+\frac{\left(n^4+9 n^2-22\right) \left(n^2-1\right)^3}{207360 c n^7}\nn\\
&&-\frac{(n-2) (n-1) (n+1) (59 n^6+136 n^5+3191 n^4+6640 n^3-7279 n^2-12536 n-7491)}{21772800 c^2 n^7}\nn\\
&&+\frac{(n-3) (n-2) (n-1) (n+1) (n+2) (n+3) \left(n^2+119\right)}{453600 c^3 n^7}+b_{\mT_4\mT_4}
\eea
Squaring the above vacuum block we find
\bea
&&\Delta S_n(\psi,x)=\frac{(1+n)c}{12n\pi}(2x/\lambda_T)^2-\frac{(1+n)c}{120 n\pi^2}\frac{(n^2+11)}{12n^2}(2x/\lambda_T)^4\nn\\
&&+\frac{(1+n)c}{630n\pi^3}\frac{(4-n^2)(n^2+47)}{144n^4}(2x/\lambda_T)^6-\frac{ (1+n)c}{2800 n\pi^4} (2x/\lambda_T)^8 s_8(n,c)+\cdots\nn\\
&&s_8(n,c)=\frac{88 (n^2-9) (n^2-4) \left(n^2+119\right)+c \left(-13 n^6+1647 n^4-33927 n^2+58213\right)}{5184 (5 c+22) n^6}.\nn
\eea
The entanglement entropy is 
\bea\label{eighthent}
\Delta S_1(\psi,x)=&&\frac{c }{6 \pi }(2x/\lambda_T)^2-\frac{c }{60 \pi ^2}(2x/\lambda_T)^4+\frac{c}{315 \pi ^3}(2x/\lambda_T)^6\nn\\
&&-\frac{c}{1400\pi ^4}(2x/\lambda_T)^8\lb 1+\frac{242}{9(5c+22)}\rb+\cdots 
\eea
Note again that up to the order $(l/\lambda_T)^6$ all the contributions to the entanglement entropy come from $T$ and $T_i T_j$ and $T_i T_j T_k$. That is because $b_{\mT_{2k}}\sim (n-1)^k$ and $b_{T \mT_4}\sim (n-1)^2$. Therefore, up to this order the one-point function of $\la\psi|\mT_4|\psi\ra$ does not appear. However, at the eighth order in $l/\lambda_T$ there is a term in $b_{\mT_4\mT_4}$ and $b_{\mT_4 T T}$ that are proportional to the first power of $(n-1)$ and hence contribute to the entanglement entropy.



\section{Failure of perturbation theory for GGE}\label{nonperturb}
In this appendix, we expand the GGE in small KdV chemical potential in a perturbative expansion.
We show that demanding that the one-point functions of GGE to match those of the eigenstate is inconsistent in perturbation theory. All orders of chemical potential contribute to the one-first correction in $1/c$, and one needs a non-pertubative expression for one-point functions of GGE to compare with the eigenstate. 
We choose the following simplifying notation
\bea
&&\frac{1}{Z}tr\lb e^{-\beta H} A\rb=\la A\ra_{\beta}\nn\\
&&\frac{1}{Z}tr\lb e^{-\beta H-\mu_i Q_i} A\rb=\la A\ra_{\beta,\mu_i}\nn\\
&&\tilde{A}=A-\la A\ra_\beta
\eea
where repeated indices are summed over. Then, assuming a perturbative expansion for the GGE we have
\bea
\la A\ra_{\beta,\mu_i}=\la A\ra_\beta-\mu_i\la \tilde{A}\:\tilde{Q}_i\ra_\beta+\frac{\mu_i\mu_j}{2} \la \tilde{A}\tilde{Q}_i\tilde{Q}_j\ra_\beta+O(\mu_i\mu_j\mu_k)
\eea
Taking $A$ to be the KdV current $J_{2k}$ we have
\bea\label{pertex}
&&\la T\ra_{\beta,\mu_i}=\la T\ra_\beta-\mu_i\la \tilde{T}\:\tilde{Q}_i\ra_\beta+\frac{\mu_i\mu_j}{2} \la \tilde{T}\tilde{Q}_i\tilde{Q}_j\ra_\beta+O(\mu_i\mu_j\mu_k)\nn\\
&&\la J_{2k}\ra_{\beta,\mu_i}=\la J_{2k}\ra_\beta-\mu_i\la \tilde{J}_{2k}\:\tilde{Q}_i\ra_\beta+\frac{\mu_i\mu_j}{2} \la \tilde{J}_{2k}\tilde{Q}_i\tilde{Q}_j\ra_\beta+O(\mu_i\mu_j\mu_k)
\eea

In (\ref{pertex}) it is understood that the index $i=2m-1$ is summed over, and $m$ runs over $2$ to $\infty$.
The first term in the series above $\la J_{2k}\ra_\beta\sim c^k$ at large $c$. 
The above expansion is a valid perturbation theory if chemical potentials are suppressed at large $c$ by $\mu_{2m-1}\sim c^{-\alpha(m)}$.
Since the disconnected piece of $\la \tilde{J}_{2k}\tilde{Q}_{2m-1}\ra_\beta$ is zero, at large central charge $\la \tilde{J}_{2k}\tilde{Q}_{2m-1}\ra_\beta=O(c^{k+m-1})$.
The first order term gives us the condition $\alpha(m)>m-1$, and from the second order term we find $\alpha(m)> m$.

In order to match this with the energy eigenstate we should solve for $\mu_i$ such that 
\bea
\la T\ra_{\beta,\mu_i}^k=\la J_{2k}\ra_{\beta,\mu_i}.
\eea
If $\mu_i$ are suppressed by powers of $c$, we can try to impose the above condition by setting
\bea\label{firstorder}
\sum_{m=2}^\infty\mu_{2m-1}\lb k \la T\ra_\beta^{k-1}\la \tilde{T}\tilde{Q}_{2m-1}\ra_\beta-\la \tilde{J}_{2k}\tilde{Q}_{2m-1}\ra_\beta \rb=\la J_{2k}\ra_\beta- \la T\ra_{\beta}^k+O(c^{k-2})
\eea
The coefficient of $\mu_{2m-1}$ in the left hand side of (\ref{firstorder}) is $O(c^{k+m-1})$, hence the each term in the sum on the left is 
scales at bet as $c^{k-1}$; while on the right hand side we have terms that are order $c^{k-1}$.
The only option is to take $\alpha=m$. According to the perturbation expansion (\ref{pertex}) this means that the higher orders terms in $\mu$ contribute to the same order in $c$. 
In order to make sense of the perturbation theory we should be able to truncate the sum on the left to a finite number of terms. Say we keep the coefficients $\mu_{2m-1}\sim c^{-\alpha(m)}$ with for $\alpha(m)=m$ for $m\leq C$ and $\alpha(m)<m$ for $m>C$, where $C$ is a finite number. Then, we have $C$ unknowns ($\mu_{2m-1}$ for $m\geq C$) that should satisfy an infinite number of equations at the firt order in $1/c$ in (\ref{firstorder}). We take this over-constrained system of equations as an indication that the question of finding a GGE with the same one-point functions as the energy eigenstate is non-perturbative in nature.

%
Below, we develop the perturbation theory in small chemical potential further, even though it does not shed light on our study of ETH.
In the remainder of this appendix, we compute some of the one-point function of $J_4$ and $T$ in an example of a GGE with only $\mu_3$ turned on.
The conserved currents are $T(\omega)$ and $(T T)(\omega)=\mT_4(\omega)+\frac{3}{10}\p_\omega^2 T(\omega)$ on the thermal cylinder of circumference $\beta$. Under a conformal transformation $z=f(\omega)$ the currents change according to
\bea
&&T(\omega)=f'^2T(f)+\frac{c}{12}Schw(f)\nn\\
&&(T T)(\omega)=\mT_4(\omega)+\frac{3}{10}\p_\omega^2 T(\omega)\nn\\
&&=f'^4 \mT(f)+\frac{(5c+22)}{30}Schw(f)\lb f'^2 T(f)+\frac{c}{24}Schw(f)\rb\nn\\
&&+\frac{3}{10}\p^2_\omega \lb f'^2 T(f)+\frac{c}{12}Schw(f)\rb\nn\\
&&Schw(f)=\frac{f'''}{f'}-\frac{3}{2}\lb\frac{f''}{f'} \rb^2.
\eea
Mapping the thermal cylinder to the complex plane by $z=e^{2\pi \omega/\beta}$ we find (see \cite{Gaberdiel:1994fs})
\bea
&&T(\omega)=\lb \frac{2\pi}{\beta}\rb^2\lb z^2 T-\frac{c}{24}\rb\nn\\
&&(T T)(\omega)=\lb \frac{2\pi}{\beta}\rb^4 \lb z^4 \mT_4(z)+\mathcal{D}_2 T(z)+\frac{c(5c+22)}{2880}\rb\nn\\
&&\mathcal{D}_2=\frac{3}{10}\lb z^4 \p^2+5z^3\p -\frac{5(c-10)}{18}z^2\rb.
\eea
From this it is immediately clear that on the complex plane
\bea
&&\tilde{T}(z)=\lb \frac{2\pi}{\beta}\rb^2\: z^2 T(z)\nn\\
&&\tilde{J}_4(z)=\lb \frac{2\pi}{\beta}\rb^4\lb z^4 \mT_4(z)+\mathcal{D}_2 T(z)\rb.
\eea
After some straightforward algebra we find
\bea
&&\la \tilde{T}(0)\tilde{Q}_3\ra_\beta=\lb \frac{2\pi}{\beta}\rb^3\int_0^\infty\frac{dz}{z}\la T(-1) \lb z^4 \mT_4(z)+\mathcal{D}_2 T(z)\rb\ra=-\lb \frac{2\pi}{\beta}\rb^3\frac{c(5c+22)}{720}\nn\\
&&\la \tilde{T}(0)\tilde{Q}_3\tilde{Q}_3\ra_\beta=\lb \frac{2\pi}{\beta}\rb^6\int_0^\infty\frac{dzdz'}{z z'}\la T(-1) \lb z^4 \mT_4(z)+\mathcal{D}_2 T(z)\rb\lb z'^4 \mT_4(z')+\mathcal{D}_2 T(z')\rb\ra\nn\\
&&=\lb \frac{2\pi}{\beta}\rb^6\frac{c(5c+22)(7c+74)}{8640}.
\eea
and for the KdV current
\bea
&&\la \tilde{J}_4(0)\tilde{Q}_3\ra_\beta=\lb \frac{2\pi}{\beta}\rb^3\frac{c(5c+22)(7c+74)}{60480}\nn\\
&&\la \tilde{J}_4(0)\tilde{Q}_3\tilde{Q}_3\ra_\beta=\lb\frac{2\pi}{\beta}\rb^6 \frac{c(5c+22)}{10}\lb \lb \frac{5c+22}{360}\rb^2+\frac{(5c+43)}{300}\rb
\eea
Here, we have used the following three-point functions
\bea
&&\la T(\infty) T(1) T(0)\ra=c, \qquad \la T(\infty) T(1) \mT_4(0)\ra=\frac{c(5c+22)}{10}\nn\\
&&\la \mT_4(\infty) T(1)\mT_4(0)\ra=\frac{2c(5c+22)}{5},\qquad \la \mT_4(\infty) \mT_4(1)\mT_4(0)\ra=\frac{c(5c+22)(5c+64)}{25}.\nn
\eea

After some algebra we find that the expectation value of currents in the GGE in the small chemical potential limit is given by 
\bea\label{pert1pt}
&&tr(\rho_{\beta,\mu} T(0))=\lb \frac{2\pi}{\beta}\rb^2\lb -\frac{c}{24}+\frac{(2\pi)^3 \mu_3}{\beta^3}\frac{c(5c+22)}{720}\rb\nn\\
&&+\frac{\mu^2}{2}\lb \frac{2\pi}{\beta}\rb^6 \lb\frac{c(5c+22)(7c+74)}{8640}\rb+O(\mu^3/\beta^9)\nn\\
&&tr(\rho_{\beta,\mu} (T T)(0))=
\lb \frac{2\pi}{\beta}\rb^4\lb \frac{c(5c+22)}{2880}-\frac{(2\pi)^3 \mu}{\beta^3}\frac{c(5c+22)(7c+74)}{60480}\rb\nn\\
&&+\frac{\mu^2}{2}\lb \frac{2\pi}{\beta}\rb^6 \frac{c(5c+22)}{10}\lb \lb \frac{5c+22}{360}\rb^2+\frac{(5c+43)}{300}\rb+O(\mu^3/\beta^9).
\eea
From which we obtain 
\bea\label{chargesGGE}
&&tr(\rho_{GGE}H)=L\lb \frac{2\pi}{\beta}\rb^2\lb\frac{c}{12}-\frac{(2\pi)^3 \mu_3}{\beta^3}\frac{c(5c+22)}{360}\rb\nn\\
&&+\frac{\mu^2}{2}\lb \frac{2\pi}{\beta}\rb^6 \lb\frac{c(5c+22)(7c+74)}{4320}\rb+O(\mu^3/\beta^9)\nn\\
&&tr(\rho_{GGE}Q_{3})=L\lb \frac{2\pi}{\beta}\rb^4\lb \frac{c(5c+22)}{2880}-\frac{(2\pi)^3 \mu}{\beta^3}\frac{c(5c+22)(7c+74)}{60480}\rb\nn\\
&&+\frac{\mu^2}{2}\lb \frac{2\pi}{\beta}\rb^6 \frac{c(5c+22)}{10}\lb \lb \frac{5c+22}{360}\rb^2+\frac{(5c+43)}{300}\rb+O(\mu^3/\beta^9)
\eea
where we have suppressed the $\mu^3/\beta^9$ corrections.


\bibliographystyle{utphys}

\bibliography{ETHCFT}

\end{document}